\documentclass[sigconf]{acmart}

\AtBeginDocument{%
  \providecommand\BibTeX{{%
    \normalfont B\kern-0.5em{\scshape i\kern-0.25em b}\kern-0.8em\TeX}}}

\copyrightyear{2024}
\acmYear{2024}
\setcopyright{rightsretained}
\acmConference[UIST '24]{The 37th Annual ACM Symposium on User Interface Software and Technology}{October 13--16, 2024}{Pittsburgh, PA, USA}
\acmBooktitle{The 37th Annual ACM Symposium on User Interface Software and Technology (UIST '24), October 13--16, 2024, Pittsburgh, PA, USA}
\acmDOI{10.1145/3654777.3676415}
\acmISBN{979-8-4007-0628-8/24/10}

\usepackage{xspace}
\usepackage{subfig}
\usepackage{url}
\usepackage{xcolor}

\begin{document}

\title[Computational Trichromacy Reconstruction]{Computational Trichromacy Reconstruction: \\ Empowering the Color-Vision Deficient to Recognize Colors \\ Using Augmented Reality\vspace{10pt}}

\author{Yuhao Zhu}
\affiliation{%
  \institution{University of Rochester}
  \city{Rochester}
  \state{NY}
  \country{USA}
}
\orcid{0000-0002-2802-0578}
\email{yzhu@rochester.edu}

\author{Ethan Chen}
\affiliation{%
  \institution{University of Rochester}
  \city{Rochester}
  \state{NY}
  \country{USA}
}
\orcid{0009-0008-1250-769X}
\email{echen48@ur.rochester.edu}

\author{Colin Hascup}
\affiliation{%
  \institution{University of Rochester}
  \city{Rochester}
  \state{NY}
  \country{USA}
}
\orcid{0009-0007-6266-846X}
\email{chascup@u.rochester.edu}

\author{Yukang Yan}
\affiliation{%
  \institution{University of Rochester}
  \city{Rochester}
  \state{NY}
  \country{USA}
}
\orcid{0000-0001-7515-3755}
\email{yukang.yan@rochester.edu}

\author{Gaurav Sharma}
\affiliation{%
  \institution{University of Rochester}
  \city{Rochester}
  \state{NY}
  \country{USA}
}
\orcid{0000-0001-9735-9519}
\email{gaurav.sharma@rochester.edu}

\renewcommand{\shortauthors}{Zhu, et al.}

\begin{abstract}
We propose an assistive technology that helps individuals with Color Vision Deficiencies (CVD) to recognize/name colors.
A dichromat's color perception is a reduced two-dimensional (2D) subset of a normal
trichromat's three dimensional color (3D) perception, leading to confusion when visual stimuli that appear identical to the dichromat are referred to by different color names.
Using our proposed system, CVD individuals can interactively induce distinct perceptual changes to originally confusing colors via a computational color space transformation.
By combining their original 2D precepts for colors with the discriminative changes, a three dimensional color space is reconstructed, where the dichromat can learn to resolve color name confusions and accurately recognize colors.
Our system is implemented as an Augmented Reality (AR) interface on smartphones, where users interactively control the rotation through swipe gestures and observe the induced color shifts in the camera view or in a displayed image. Through psychophysical experiments and a longitudinal user study, we demonstrate that such rotational color shifts have discriminative power (initially confusing colors become distinct under rotation) and exhibit structured perceptual shifts dichromats can learn with modest training. The AR App is also evaluated in two real-world scenarios (building with lego blocks and interpreting artistic works); users all report positive experience in using the App to recognize object colors that they otherwise could not.
\end{abstract}

\ccsdesc[500]{Human-centered computing~Accessibility technologies}
\ccsdesc[300]{Human-centered computing~Information visualization}

\keywords{Assistive Technology, Color Vision Deficiency, Dichromat, Augmented Reality, Color Recognition, Color Naming}

\newcommand{\website}[1]{{\tt #1}}
\newcommand{\program}[1]{{\tt #1}}
\newcommand{\benchmark}[1]{{\it #1}}
\newcommand{\fixme}[1]{{\textcolor{red}{\textit{#1}}}}

\newcommand*\circled[2]{\tikz[baseline=(char.base)]{
            \node[shape=circle,fill=black,inner sep=1pt] (char) {\textcolor{#1}{{\footnotesize #2}}};}}

\ifx\figurename\undefined \def\figurename{Figure}\fi
\renewcommand{\figurename}{Fig.}
\renewcommand{\paragraph}[1]{\textbf{#1} }
\newcommand{\figline}{{\vspace*{.05in}\hline}}

\newcommand{\Sect}[1]{Section~\ref{#1}}
\newcommand{\Fig}[1]{Figure~\ref{#1}}
\newcommand{\Tbl}[1]{Table~\ref{#1}}
\newcommand{\Eqn}[1]{Equation~\ref{#1}}
\newcommand{\Apx}[1]{Appendix~\ref{#1}}
\newcommand{\Alg}[1]{Algorithm~\ref{#1}}

\newcommand{\specialcell}[2][c]{\begin{tabular}[#1]{@{}c@{}}#2\end{tabular}}
\newcommand{\note}[1]{\textcolor{red}{#1}}

\newcommand{\proj}{\textsc{Cicero}\xspace}
\newcommand{\algo}{\textsc{SpaRW}\xspace}
\newcommand{\mode}[1]{\underline{\textsc{#1}}\xspace}
\newcommand{\sys}[1]{\underline{\textsc{#1}}}

\newcommand{\no}[1]{#1}
\renewcommand{\no}[1]{}
\newcommand{\RNum}[1]{\uppercase\expandafter{\romannumeral #1\relax}}

\newcommand{\change}[1]{#1}

\def\cG{{\mathcal{G}}}
\def\cF{{\mathcal{F}}}
\def\cI{{\mathcal{I}}}
\def\cN{{\mathcal{N}}}
\def\bh{{\mathbf{h}}}
\def\bp{{\mathbf{P}}}
\def\bq{{\mathbf{Q}}}

\def\p{Protanope\xspace}
\def\d{Deuteranope\xspace}
\def\t{Tritanope\xspace}

\graphicspath{{figs/}}

\begin{teaserfigure}
  \includegraphics[width=\textwidth]{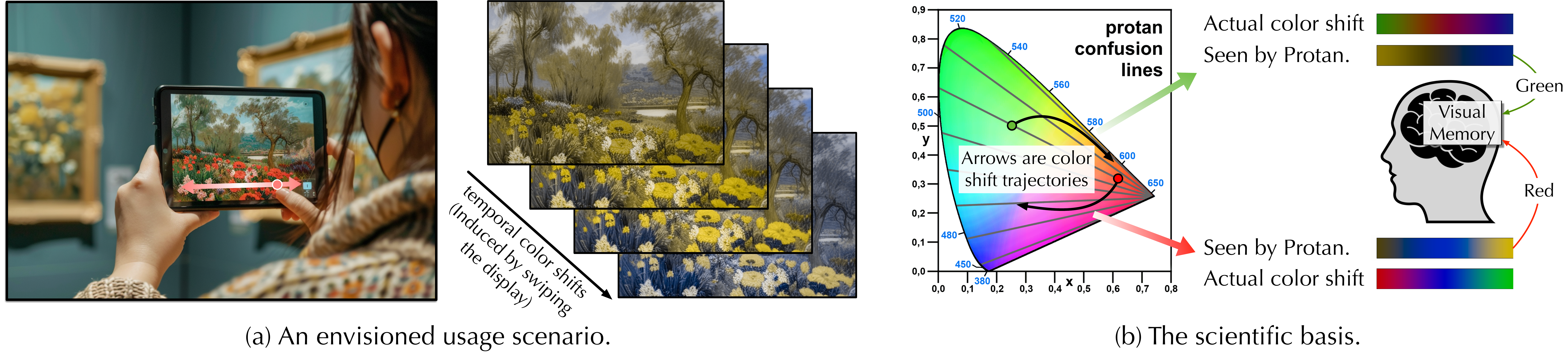}
  \caption{(a) An envisioned usage scenario where a Protanope (individual with a kind of red-green color blindness) views a colored painting in a museum using our mobile App by swiping her finger on the screen. The swiping motion, causes different temporal perceptual shifts in initially confusing colors (e.g., green vs. red here). These shifts augment the user's perception allowing her to recognize/name the colors in the painting, based on prior learning of the associations. (b) The scientific basis of our work: controlled by the user's swiping, we computationally induce a temporal shift in the colors that produces distinct perceptual changes in colors that are initially confusing to a dichromat. Users learn to associate the name of a color with the initial appearance and the discriminative shift, and later can use these to resolve names of otherwise confusing colors. The CIE 1931 xy chromaticity diagram~\cite{confusion_lines_xy} is annotated with Protan ``confusion lines'': colors on the same confusion line are indistinguishable for a Protanope, e.g., the green and red indicated by the hollow circle markers in the diagram. Upon swiping, the two confusing colors undergo different color shifts: the green shifts toward yellow and the red shifts toward blue.
  }
  \Description{The scientific basis of our work and an envisioned usage scenario.}
  \label{fig:teaser}
\end{teaserfigure}

\maketitle

\section{Introduction}
\label{sec:intro}

Approximately 13 million individuals in the U.S. and 350 million worldwide have some form of color vision deficiency (CVD)~\cite{Birch:GreenRedWorldStatistics:JOSAA2012}.
Individuals with CVD have restricted career options, limited driving rights in certain countries, and are less confident in social interactions.
In a 2020 survey, one third of the students with CVD indicated that color blindness affected their confidence in school and, before they learned of their color vision deficit, 30\% felt they might be ``slow learners''~\cite{cvd_report}.
CVD is rising in geographic areas that have been settled by incoming migrants where people are more likely to have mixed race genes in their genetic history~\cite{Birch:GreenRedWorldStatistics:JOSAA2012}.

Many assistive technologies for CVD exist.
The vast majority focus on color \textit{discrimination}, i.e., being able to distinguish confusing colors for a CVD individual, by applying a transformation (``color filter'') such that initially confusing colors become different~\cite{ribeiro2019recoloring}.
Color discrimination alone, however, is insufficient in many real-world scenarios where we have to \textit{recognize} (name) an object's color (``pass me that pink block'' or ``look at the person in red shirt'').

Color names are developed in the context of trichromatic vision, where colors are encoded in a 3D space.
Dichromats miss the function of a retinal photoreceptor type and, thus, have a reduced color vision where colors are encoded in a 2D space for them.
Therefore, there are colors that appear distinct to a trichromat (and, thus, have different assigned names) but appear identical to a dichromat.
This creates significant confusions for dichromats to name colors.

To assist CVD individuals in color recognition, we develop a smartphone-based Augmented Reality (AR) application.
The idea is illustrated in \Fig{fig:teaser} and detailed in \Sect{sec:idea}.
The key is to augment the 2D color percept of a dichromat with an \textit{additional dimension} of information such that colors are, once again, encoded in a 3D space for dichromats.
The additional dimension is induced via the temporal modulation of colors: as a user swipes the finger in the App, we apply a color-space transformation such that originally confusing colors undergo distinct color shifts.
The combination of the initial 2D color precept with the induced temporal shifts reconstructs a new 3D space for the user.

Critically, this new 3D space empowers users to \textit{learn} to recognize colors.
By spending time interacting with our system, users build an intuition of how originally confusing colors undergo distinct temporal color shifts.
Users then learn to associate different color names with different shifts, thereby recognizing colors.

We conduct two studies to demonstrate the effectiveness of temporal color shifts.
We first perform extensive psychophysics (on 16 CVD individuals with almost 100 total hours of study) to show that rotational color shifts have discriminative power, i.e., initially confusing colors become distinct upon rotation.
Leveraging the color shifts, participants' color discrimination improves across different CVD types, and the results are statistically significant (\Sect{sec:pp}).

We then conduct a nine-day longitudinal study on eight CVD individuals (\Sect{sec:long}).
Participants first trained themselves to recognize four pairs of confusing colors by associating the color shift patterns with color names.
Over the next few days participants are asked to recognize new colors that they have not seen during training.
The recall performance is significantly above the chance level, indicating that the shift patterns are easy to learn and that users can generalize what they have learned from training to recognize new colors over an extended period of time.

We evaluate the effectiveness of our method and the AR interface on two real-world scenarios, building with Lego blocks and observing artistic images, both requiring identifying objects by color (\Sect{sec:real}).
We provide an empirical, experiential report on the two real-world studies.
Overall, all participants report positive experience, where by using our AR App they are able to recognize colors that they otherwise could not accurately name.

To inform our design choices and objectives for this project, we closely engaged with two CVD individuals (distinct from the 16 study participants), who participated in our weekly meetings and provided feedback/suggestions via open-ended dialogues.
This is in line with the best practices suggested by a recent study on CVD individuals' experience with assistive technologies~\cite{geddes202330}.
The protocols for our experiments involving human subjects were approved by our Internal Review Board (IRB).

In summary, the paper makes the following contributions:
\begin{itemize}
        \item We propose a framework for reconstructing trichromatic color perception for CVD individuals.
        The idea is to augment their native (unaided) 2D color precept with a third dimension.
        The third dimension is interactively-induced by users to shift originally confusing colors in a color space.
        The perceptual shifts are distinct for confusing colors and, through learning,
        enable users to resolve confusions and accurately recognize colors.
        \item We develop a practical realization of our framework using rotational color shifts about the gray-axis in RGB space, which allows computationally efficient implementation and enable real-time interactivity.
        \item Through psychophysical experiments, we demonstrate that the rotational shifts have discriminative power and that users can indeed learn the patterns of induced shifts to resolve color confusions and accurately recognize colors.
        \item We implement our proposed system as smartphone AR App. Through user studies, we show the feasibility of our AR App in assisting CVD individuals to recognize object colors using two real-world tasks.
\end{itemize}

\section{Preliminaries}
\label{sec:bck}

\begin{figure*}[t]
    \centering
    \subfloat[Colors $\bp$ and $\bq$ appear distinct for trichromats and are assigned different names but are perceived as identical by a \d, leading to naming confusion.
    This is because the \d's color perception is 2D, missing the M cones.
    Using our system, the user can induce a color-space transformation, $\Delta_p$ and $\Delta_q$, to $\bp$ and $\bq$, respectively.
    Critically, \d perceives $\Delta_p$ and $\Delta_q$ as distinct perceptual changes.
    The changes provide a new dimension that, when combined with the initial 2D percept of a color, positions the color in a new 3D space for a \d, eliminating naming confusion.
    The shift pattern is regular (cycling a color through different hues) and, thus, can be learned and generalizes locally.]{
      \label{fig:idea}
      \includegraphics[height=1.37in]{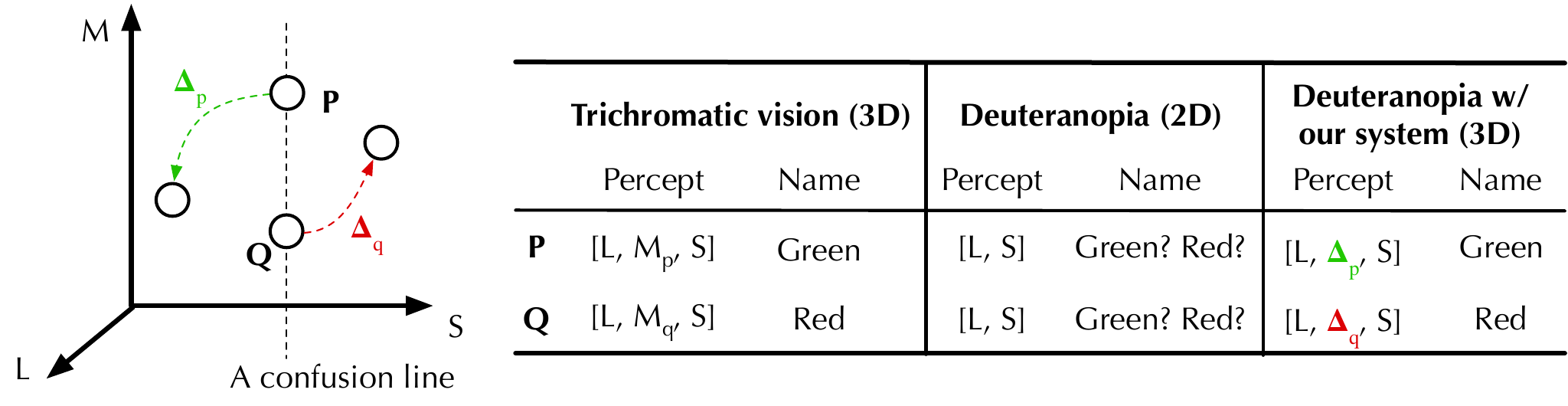}
    }
    \hspace{5pt}
    \subfloat[Deuteranopia confusion lines in the CIE 1931 xy-chromaticity diagram~\cite{confusion_lines_xy}. $\bp$ and $\bq$ lie on a confusion line.]{
      \label{fig:idea_f}
      \includegraphics[height=1.37in]{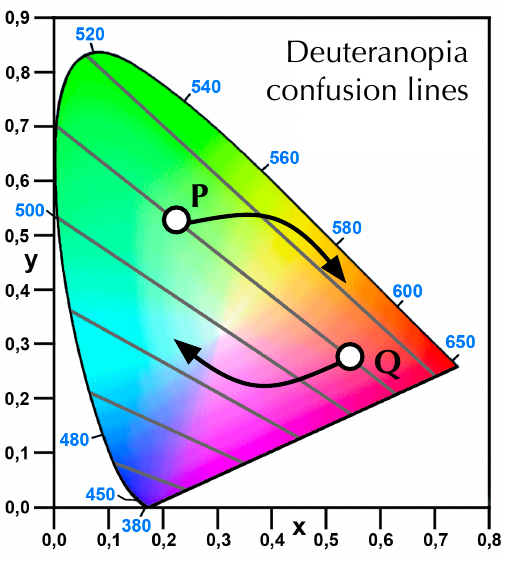}
    }
  \caption{Illustration demonstrating color naming confusion for CVD individuals and how our system resolves such confusion.
  The xy-chromaticity space in \protect\subref{fig:idea_f} can be obtained by a $3\times3$ linear transformation of the 3D LMS space in \protect\subref{fig:idea} followed by a perspective projection.
  The xy diagram is commonly used in color science, because it conveniently allows us to visualize colors in a 2D representation~\cite{sharma2017digital, chromaticitytutorial}.
  Note that the Deuteranopia confusion lines, which are parallel to the M-axis in the LMS space, are now converging in the xy space.}
    \label{fig:idea}
\end{figure*}

\paragraph{Color Vision Basics.}
Human color perception is trichromatic because humans have three types of cone photoreceptors, referred to as the long (L), medium (M), and short (S) wavelength sensitive cones. Retinal sensations that (eventually) encode colors can be geometrically represented as points in the 3D LMS space where the three L, M, and S, axes correspond to the responses of the respective cones.  For instance, the two colors $\bp$ and $\bq$ in \Fig{fig:idea} are represented as [L, M$_p$, S] and [L, M$_q$, S], respectively, in the LMS space. These two colors are distinct for normal trichromats, as they excite different combinations of the cone responses.
To a fair degree of accuracy, commonly used RGB color spaces can be obtained by applying a $3 \times 3$ linear transformation to the LMS space, followed by an channel-wise R, G, B nonlinearity~\cite{sharma2017digital}.

\paragraph{CVD.}
Dichromatic individuals see colors only in a 2D space, because they lack (the functionality of) one cone type.
For instance, Deuteranopes lack the M cones and, thus, any color is encoded only in the L and S cone excitations, resulting in a 2D color space.
Therefore, any colors that differ \textit{only} in the M dimension are seen as the same color by a \d.
For instance, $\bp$ and $\bq$ in \Fig{fig:idea}, distinct to trichromats, are seen as the same color by a \d.

A line parallel to the M dimension in the LMS space is called a Deuteran ``\textbf{confusion line}'', as all colors on that line look the same to a \d; we call such colors ``confusion colors''.
\Fig{fig:idea} plots only one such line that connects $\bp$ and $\bq$, although it is easy to see that there are infinitely many confusion lines.

The confusion lines are more commonly visualized in the CIE 1931 \textbf{xy-chromaticity} diagram~\cite{sharma2017digital, chromaticitytutorial}, as illustrated in \Fig{fig:idea_f}.
The xy space is a perspective projection from the LMS space that provides a useful 2D representation of colors by discarding the luminance dimension that is indicative of brightness/lightness.
We will primarily rely on the xy diagram for the rest of the paper.

There are three forms of dichromatic vision, based on the missing cone type.
While Deuteranopes lack M cones, Protanopes and Tritanopes lack L cones and S cones, respectively.
As a result, Deuteran, Protan, and Tritan confusion lines are parallel to the M, L and S axis in the LMS space, respectively.

In addition to strict dichromatic vision, another important form of CVD is anomalous trichromatic vision, where individuals have all three cone types, but the sensitivity of a cone type deviates from normal.
Protanomaly, Deuteranomaly, and Tritanomaly are the names given to the three types of anomalous trichromatic vision.
The ability to discriminate colors is stronger for anomalous trichromats than their dichromatic counterparts.
For instance, it is shown that, from a modeling perspective, not all colors on a Deuteranopia confusion line are confusing to Deuteranomalous individuals; only colors within a small segment are~\cite{Flatla:ColourID:CHI2015}.

\paragraph{Color Discrimination vs. Recognition.}
Color discrimination refers to the ability to tell the \textit{difference} between two colors, whereas color recognition requires a stronger capability: to assign a \textit{name} to a color.
Recognition requires uniquely associating a given color percept to its name in a common vocabulary.
Normal trichromats, during their early cognitive development, are told names of the colors of objects.
They then implicitly build a mapping between their 3D precept for a color and the corresponding color name. Conceptually, a trichromat's visual memory stores such a mapping from 3D color precepts to color names, which they can later retrieve to name colors they encounter.

Such recognition tasks are harder for CVD individuals, because of their reduced 2D perception.
For instance in \Fig{fig:idea}, lights that produce the distinct responses $\bp$ and $\bq$ a trichromat are assigned different color names (e.g., red and green),
but they have the same [L, S] cone responses in a \d's vision and, thus, cannot be named accurately by a \d. 

\paragraph{Color Names.}
Assigning names to colors is a research area in its own right~\cite{heer2012color}.
Many color name dictionaries exist, such as the Unix X11 Colors in \texttt{/usr/lib/X11/rgb.txt} (752 names), HTML 4.01 Basic Colors (16 names)~\cite{web401colors}, CSS Colors (140 names)~\cite{csscolors}, and NTC Colors (1566 names)~\cite{ntcjs}.
To balance the trade-off between coarse color naming (with too small of a dictionary) and using obscure names that are uncommon in everyday vocabulary (with too big of a dictionary), we use the color dictionary from the IEEE PWG 5101.1 standard~\cite{pwg}, which has 20 names, each with two additional dark and light variants (except black which has only one variant and white which has no variant), totaling 57 names.

Any given color is named based on its closest color in our dictionary according to the CIELAB $\Delta$E (1976) metric, a commonly used color different metric that emphasizes perceptual uniformity~\cite{sharma2017digital}.

\section{Related Work}
\label{sec:related}

A wide range of assistive technologies have been proposed for CVD. We discuss relevant related work in four groups: (1) computational approaches that aid in color discrimination (\Sect{sec:related:cd}), (2) methods that assist with color recognition (\Sect{sec:related:cr}), (3) optical assistive technologies (\Sect{sec:related:op}), and (4) sensory substitution techniques (\Sect{sec:related:sen}).

\subsection{Computational Color Discrimination}
\label{sec:related:cd}

Most existing technologies aim to help individuals with CVD to discriminate confusing colors by using computational ``re-coloring'' algorithms. The idea is to apply a transformation to colors in an image (colloquially referred to as color filters) so that initially confusing others become distinct. We discuss a few of the key approaches below, referring interested readers to Ribeiro and Gomes~\cite{ribeiro2019recoloring} and Geddes et al.~\cite{geddes202330} for two comprehensive surveys of the techniques and their effectiveness in the wild. A recent technical report~\cite{Hung:EhnForColorDef:cie240_2020} from the International Commission on Illumination (CIE) also categorizes and summarizes some key approaches.

Considerable research effort has been devoted to recoloring algorithms optimized for particular metrics of interest, such as enhancing contrast~\cite{jefferson2007interface, jefferson2006accommodating, flatla2013sprweb, flatla2012ssmrecolor, wakita2005smartcolor, rasche2005detail, oshima2016modeling, nunez2018optimizing, machado2010real}, maintaining color consistency~\cite{jefferson2007interface, chen2011efficient, lee2011adaptive, oshima2016modeling, nunez2018optimizing}, and preserving naturalness of the image~\cite{flatla2013sprweb, flatla2012ssmrecolor, anagnostopoulos2007intelligent, oshima2016modeling, nunez2018optimizing}.
For the optical see-through augmented reality setting, ChromaGlasses~\cite{langlotz2018chromaglasses} and the subsequent work~\cite{sutton2022seeing} have proposed selective modification of identified critical colors in the scene by using a semi-transparent display aligned with pixel precision to the scene content.
Hasana et al.~\cite{hasana2019improving} uses temporal modulation, similar to our system, but demonstrates the feasibility only for discrimination.

Many re-coloring tools have been built into widely used platforms, such as the Color Enhancer Chrome extension provided by Google~\cite{colorenhancer2}, the CVD filters built into iOS~\cite{iosacc} and Android~\cite{androidacc}, and the color filters in Windows~\cite{WindowsColorFiler2022}.

\paragraph{Distinctions.}
Recoloring methods pose several limitations that are summarized in a recent user study~\cite{geddes202330}. First, they present a modified rendering that is often completely divorced from the native perception of the user, which can be unnatural and disorienting. A consistent finding in surveys is that users favor ``on-demand'' interfaces that they can access as required, instead of the ``always-on'' mindset that many recoloring methods adopt.
In contrast, in our proposed approach, the default presentation is the unaltered natural view that users are already accustomed to.
By swiping, users introduce perceptual changes in confusing colors ``on-demand.''

Second, recoloring methods do not help color recognition, which is a common user complaint found in the user study~\cite{geddes202330}.
This is fundamentally because recoloring algorithms are still constrained by the dichromat's two-dimensional color perception.
In contrast, our methods uses temporal color modulation as a proxy for the missing dimension in the trichromatic color perception (induced by the swiping gesture), which we show is critical for naming confusing colors.
Critically, learning the temporal modulationis enabled by our ``on-demand'' interface, where users get to decide how to swipe and devise their own learning strategy to couple temporal changes with color names.

\subsection{Computational Color Recognition}
\label{sec:related:cr}

Another class of assistive technologies helps CVD individuals recognize/identify colors.
This is typically accomplished by introducing an overlay or local variation in viewed imagery.
They enable color recognition by using different patterns for confusing colors.
A variety of different overlays have been used including (a) different patterns for confusing colors, such as symbols, shapes and arrows~\cite{sajadi2012using, geddes2022improving, Flatla:ColourID:CHI2015, hung2013colour}, (b) actual color names in image regions tiled based on color~\cite{Flatla:ColourID:CHI2015}, and (c) highlighting user specified colors to make them more salient~\cite{Flatla:ColourID:CHI2015,tanuwidjaja2014chroma}, which has also been used in the AR setting with Google Glass~\cite{tanuwidjaja2014chroma}.

\paragraph{Distinctions.}
A main limitation of overlaying patterns over colors is that it introduces visual occlusion or distraction, which we avoid in our proposed approach.

The overlaying methods also apply a \textit{one-off} augmentation to the visual field, leaving users little control.
Thus, the number of colors that can be recognized is limited by the discrete set of patterns supplied by the system.
Our philosophy, in contrast, is to give the control back to the users: users voluntarily induce continuous perceptual changes to confusing colors in the visual field and learn to associated shift patterns with color names. While there are tools with interfaces that allow users to point to a particular location in the imagery and identify the color at that location, these are also suitable only in the limited setting where the goal is to recognize the color of a specific object or a small number of objects~\cite{Flatla:ColourID:CHI2015}.

\subsection{Optical Assistive Technologies}
\label{sec:related:op}

Optical assistive technologies use physical filters, typically provided as eyeglasses, to alter the spectral distribution of light in order to help discriminate otherwise confusing colors in the visual field.

Many commercially available glasses for CVD individuals, such as those from EnChroma~\cite{enchroma} and VINO~\cite{vino}, use identical spectral notch filters for both eyes.
The filter eliminates light from a narrow spectral band, where the human L and M cone sensitivities overlap the most, with the objective of amplifying the difference between L cone and M cone excitations. Therefore, in principle, the method can enhance color discrimination for anomalous trichromats, but provides no benefit for strict dichromats. In comparative tests, the functional effectiveness of such glasses for anomalous trichromats is also not definitive~\cite{patterson2022effects, gomez2018enchroma}.

Another class of optical technologies introduces binocular color disparity, where the stimulti are differentially altered for the two eyes. The idea was originated by James Maxwell~\cite{maxwell1857xviii} in the very early phase of the scientific study of color perception, and then later revived by Cornsweet~\cite{cornsweet2012visual}.
Maxwell conjectured that the disparity across the two eyes would essentially introduce a new dimension of perception, which would augment the existing 2D percept of a dichromat providing 3D color sensation and perception. Operating on the same principle, ColorBless~\cite{hau2015colorbless} introduces binocular luster effect for color discrimination and X-Chrome uses optical filters that are meant to be worn monocularly~\cite{siegil1981x}.

\paragraph{Distinctions.}
Our approach shares philosophical similarity with Maxwell's conjecture in that we also introduce a new dimension to augment a dichromat's 2D percept of color and seek to restore 3D color perception.
However, our approach starts with a user's normal
\footnote{
\change{
To characterize the color vision of individuals, we use the terms ``normal'' and ``CVD'' that are established in the literature and also used by authors who themselves have CVD~\cite{Flatla:ColourID:CHI2015, geddes202330, Knoblauch:MaxwellCornsweetConjectEval:JOSAA95}. We, however, acknowledge the need for more inclusive terminology.}}
perception of the world, whereas optical filters physically alter the incident lights and, thus, always present to users an altered color perception of the world.
Effectively, users have to first re-learn what colors look like with the filters and \textit{then} learn the effect of the additional dimension (binucular disparity), adding cognitive load.

For instance, when presented with a green object, a \p based on experience would know it is either green or red but would hesitate to name it, since the two colors are confusing.
In our method, the temporal modulation eliminates that confusion.
Using the optical filter, however, the initial percept of the object changes so that the user would not even be sure if the color is green or red, increasing the learning difficulty.

\subsection{Sensory Substitution}
\label{sec:related:sen}

Another class of CVD assistive technologies operate based on the idea of sensory substitution, where an intact modality, such as haptics or audition, is used to represent information in an impaired/absent sensory modality, e.g., color.
For instance, Amedi et al.~\cite{amedi2007shape} demonstrate physiological evidences for using auditory signals to deliver object shape information to blind people.

In the realm of CVD, both the haptic~\cite{carcedo2016hapticolor, nguyen2019exploring, wozniak2015chromaglove} and auditory modalities have been explored.
On the haptics side, Hapticolor~\cite{carcedo2016hapticolor} encodes color information into spatiotemporal vibrations on a wristband for color recognition and comparison;
Nguyen and Geddes~\cite{nguyen2019exploring} explored delivering vibration to a person's wrist and back.
On the auditory side, Soundview~\cite{van2003soundview} maps the three dimensions of the HSV color space to the frequency, pitch, and gain attributes of an acoustic wave.
EyeMusic~\cite{abboud2014eyemusic} uses musical notes generated by natural instruments to convey shape and color information.

\paragraph{Distinctions.}
Conceptually, our proposed method is also a form of sensory substitution, where the missing dimension in trichromatic vision is replaced by the temporal modulation.
Our approach, however, operates entirely within the visual modality, since the induced temporal changes are in the perceived colors.
Operating within the visual modality is advantageous for learning, because we start with and make use of the unaltered perception that the user is already familiar with (and where extensive experience grounds their color perception and terminology) without requiring users to learn a different modality.

Of course, sensory substitution could be used for extreme impairments, such as for blind individuals or monochromats, where the proposed approach would not be viable.

\section{Main Idea and Intuition}
\label{sec:idea}

\Sect{sec:idea:ui} introduces the main idea behind our AR system and what it looks like from a user's perspective.
\Sect{sec:idea:rot} describes the technical details behind the system, followed, in \Sect{sec:idea:why}, by intuitive explanation why our idea works .

\subsection{User Interface and Main Idea}
\label{sec:idea:ui}

\begin{figure}[t]
  \centering
  \includegraphics[width=\columnwidth]{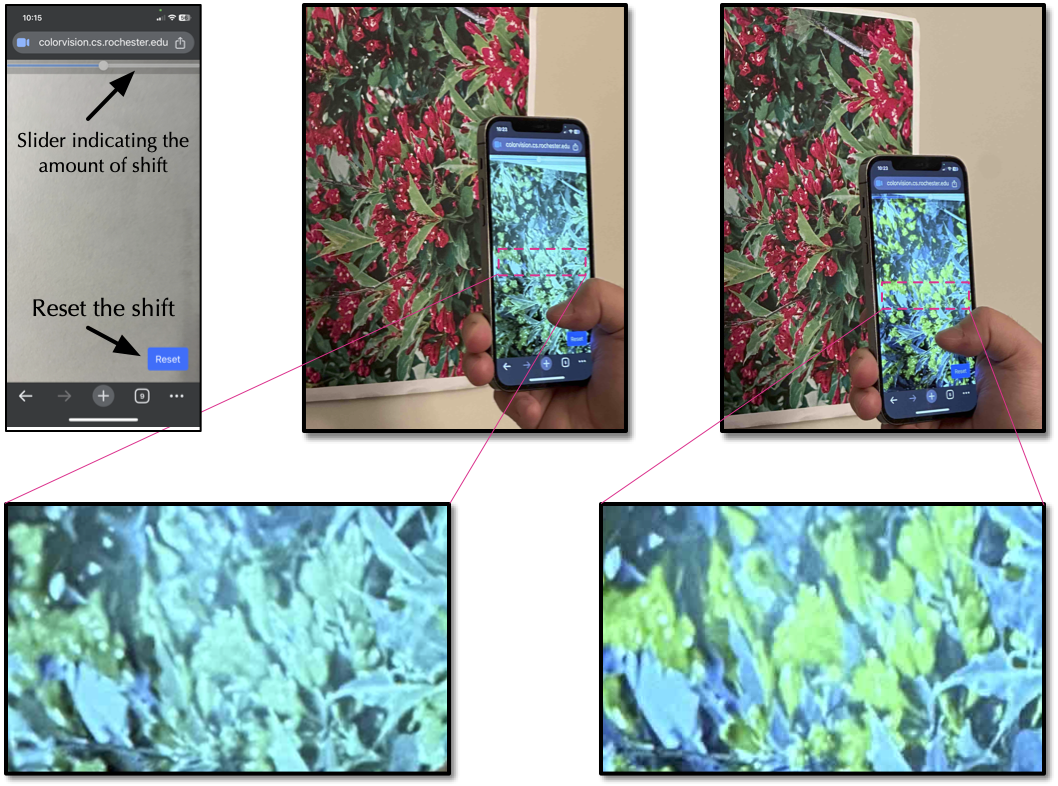}
  \caption{User swipes the finger to shifts colors of objects in the physical world.
  Initially confusing colors have different shift patterns (bottom blow-ups), which the user learns over time and uses to name/recognize colors.
  The slider at the top indicates the amount of shift applied, which our users report very useful for learning color naming.
  The button at the bottom right resets the shift to the initial colors.}
  \label{fig:demo}
\end{figure}

Our goal is to assist CVD individuals to recognize common colors (by correctly naming them).
The challenge with color naming for dichromats, as shown in \Fig{fig:idea}, is that many distinct colors share the same 2D color percept.
Our key idea is to augment the 2D percept of a dichromat with an additional dimension of information such that each color is now represented in a new 3D space, eliminating naming confusion.

\Fig{fig:demo} illustrates our AR application. The CVD user is viewing a scene through the smartphone camera.
As the user swipes her finger on the phone display, a computational color transform is applied to the image that results in distinct perceptual shifts in colors that are originally confusing to the user.
The temporal color shifts caused by the swiping introduce a new perceptual dimension that discriminates confusing colors.
The blow-ups show how the colors shift as a \d sees it (simulated).
Critically, original confusing colors, such as red flowers and green leaves, have different shift patterns.

The last table column in \Fig{fig:idea} explains our idea more formally.
While the inherent color percept of $\bp$ and $\bq$ for a \d is 2D (conceptually equivalent to the L and S cone excitation), by swiping the finger the user introduces a color transformation that shifts the points $\bp$ and $\bq$ in the trichromatic LMS space as depicted, which map into distinct perceived shifts $\Delta_p$ and $\Delta_q$ in the LS perceptual space for the \d user. The shift combined with the initial 2D percept results in a new 3D space, in which $\bp$ and $\bq$ are distinct, as they are represented by two different points in that space: [L, $\Delta_p$, S] and [L, $\Delta_q$, S].

Our hypothesis is that as a user spends more time with our system they can gradually build the intuition of how a set of originally confusing colors undergo different temporal color shifts, thereby resolving naming confusions.
For instance, with enough training a \d could learn to associate [L, $\Delta_p$, S] with green-ish color and [L, $\Delta_q$, S] with a red-ish color.
Such mappings can be later retrieved from the visual memory to name similar colors, much as trichromats associate color names with their 3D color precepts.

\change{While the discussion above uses dichromats as an example, our approach applies to both dichromats and anomalous trichromats.
Unaided, anomalous trichromats have elongated discrimination thresholds along the confusion lines. As in the case of dichromats, 
the unique temporal shifts of otherwise confusing colors can be used by anomalous trichromats to reduce the discrimination threshold and can be learned to resolve name confusion.}

\subsection{Temporal Shift via Color-Space Rotation}
\label{sec:idea:rot}

How exactly should a color shift?
There are three requirements.
First, the shift must have \textit{discriminative power}: initially confusing colors should become distinguishable at some point during the shift.
Second, the shift pattern must also be relatively easy for to CVD individuals to \textit{learn}, e.g., it should exhibit regular patterns.
Finally, underlying color transformation that produces the color shift must be \textit{computationally efficient} for real-time rendering.

With the three requirements in mind, we propose to induce the color shifts by \textit{rotating} colors about the gray-axis in the linear sRGB space.
For each camera frame encoded in the conventional nonlinear sRGB space~\cite{IEC:61966sRGBStd:1999}, we first convert all the pixels to the linear sRGB space through a computationally simple, channel-wise, 1D look-up table (a.k.a., gamma decoding).
We then perform the rotation operation.
Specifically, given a color $[r, g, b]$ expressed in the linear sRGB space, if we want to rotate it an angle $\theta$ (in radian) about the gray axis, the rotated color $[r', g', b']$ is~\cite{palazzolo1976formalism, cole2015modelling}:

\begin{align*}
	& \begin{bmatrix}
		r'\\
		g'\\
		b'
	\end{bmatrix}
	=
	\begin{bmatrix}
		c + u^2 (1-c) & u^2(1-c)-us & u^2(1-c)+us\\
		u^2(1-c)+us & c+u^2(1-c) & u^2(1-c)-us\\
		u^2(1-c)-us & u^2(1-c)+us & c+u^2(1-c)
	\end{bmatrix}
	\times
	\begin{bmatrix}
		r\\
		g\\
		b
	\end{bmatrix}\\
	& \text{where}~~~~u = 1/\sqrt{3},~~c = \cos\theta,~~s = \sin\theta
\end{align*}

Such a rotation is very computationally efficient, as it requires the same matrix multiplication for all the pixels, which can be efficiently executed on Graphics Processing Units (GPUs) on modern smartphones.
As a result, we achieve real-time (60 FPS) rendering on an iPhone 11 device.
While we did not restrict the device that our participants use, they did not encounter sluggish rendering.
We have also considered rotating in other color spaces, such as the CIE XYZ, CIELAB, CIELUV, and HSV space, which, however, would have required much more complicated transformations beyond a simply linear transformation, especially for the latter three, as they all involve non-linear transformations from the linear sRGB space.

It is worth noting that there are other methods of inducing color shifts.
The goal of the paper is to introduce color shift as a new idea for color recognition and show that rotational shifts, while simple at first glance, are sufficient for color recognition.
We discuss the design space of color shifts in \Sect{sec:disc} and leave a thorough exploration to future work.

Intuitively, such a rotation in the RGB color space shifts a color along the color wheel, changing its hue over time.
\Fig{fig:rot_rgb} and \Fig{fig:rot_xy} illustrate how the rotation looks in the RGB space and in the xy diagram, respectively, using eight colors that initially lie on a Deuteranopia confusion line.
A rotation in the RGB space leads to an elliptical trajectory in the xy-chromaticity diagram\footnote{A circle in linear RGB becomes an ellipse after a perspective projection from the linear sRGB space to the xy space.}.

Note that a color during rotation might go beyond the gamut of the sRGB color space (or for that matter, the gamut of human visual system).
In this case, we simply clip the dimension at which the color value extends beyond the gamut.
This is evident in \Fig{fig:rot_xy} where the trajectories of more saturated colors (which are closer to the gamut boundary pre-rotation) are clipped by the sRGB gamut.
\change{Empirically, we find that this clipping has little impact in practice because: 1) many colors are desaturated and do not go beyond the gamut boundary under rotation, 2) even in cases where colors go beyond the gamut boundary under rotation, users usually do not notice, because the general shift direction remains the same.}

\subsection{Why Do Rotational Shifts Work?}
\label{sec:idea:why}

\begin{figure*}[t]
    \centering
    \subfloat[We rotate colors in the linear sRGB color space about the gray axis (connecting \text{[0, 0, 0]} and \text{[1, 1, 1]}).]{
      \label{fig:rot_rgb}
      \includegraphics[width=.68\columnwidth]{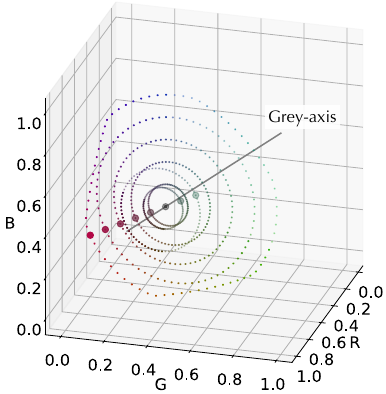}
    }
    \hspace{5pt}
    \subfloat[The same rotational shifts shown in the CIE 1931 xy-chromaticity diagram along with the sRGB gamut.]{
      \label{fig:rot_xy}
      \includegraphics[width=.68\columnwidth]{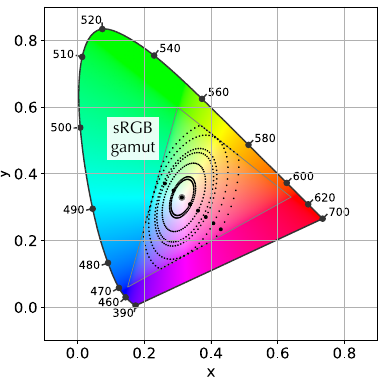}
    }
    \hspace{5pt}
    \subfloat[How different color shifts are seen by a \d in the xy-diagram~\cite{confusion_lines_xy}.]{
      \label{fig:intuition}
      \includegraphics[width=.62\columnwidth]{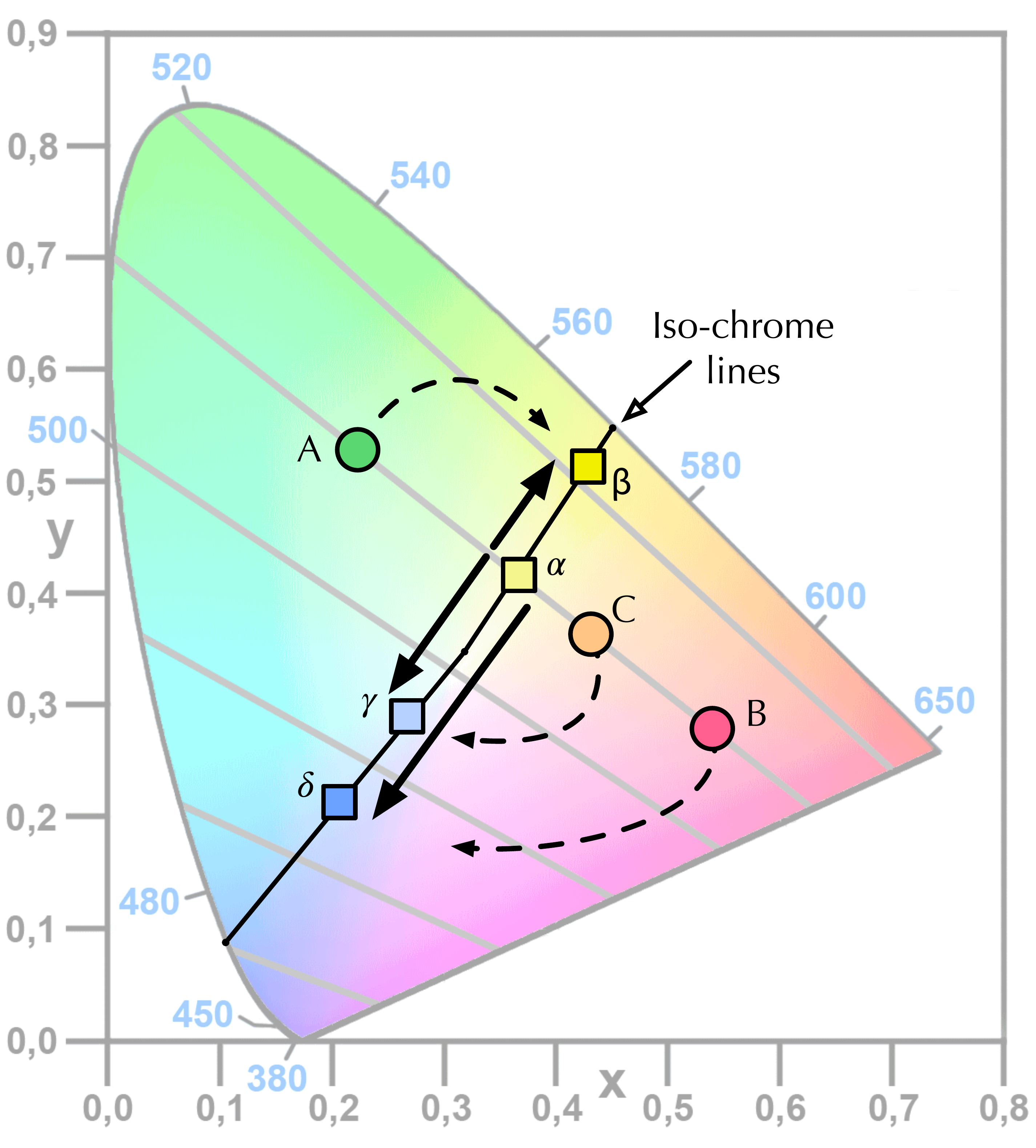}
    }
    \caption{\protect\subref{fig:rot_rgb} -- \protect\subref{fig:rot_xy}: Rotating colors about the gray axis cycles the colors through different hues.
    Notice how the sRGB gamut clips the trajectories.
    \protect\subref{fig:intuition}: Under rotational shifts, initially confusing colors are uniquely recognizable: colors on the opposite side of the iso-chrome lines (e.g., A and B) shift in opposite directions, and colors on the same side of the iso-chrome lines (e.g., B and C) shift in the same direction but with different amounts.
      The iso-chrome line contains all the colors a \d can see~\cite{brettel1997computerized, judd1948color} (disregarding luminance variations, which is inherent in the chromaticity representation),
      so the intersection of a confusion line and the iso-chrome line represents the color that a dichromat actually sees when presented with all the colors on the confusion line.
      For instance, a \d sees colors A, B, C as the color $\alpha$.
    }
    \label{fig:rotation_examples}
\end{figure*}

We now analyze the effect of our rotational color shifts.
The analysis will provide the intuition as to why rotational shifts meet the two requirements established above: rotational shifts 1) have discriminative power and 2) can be learned by CVDs.

\paragraph{Primer: Iso-chrome Lines.}
To establish the intuition, we must first understand a key question: for a set of confusion colors, what is the color that a dichromat actually sees?

Psychophysical studies~\cite{brettel1997computerized, judd1948color} suggest that all the colors that a dichromat can see lie on the so-called ``iso-chrome'' lines, which contains all the colors that a dichromat can correctly perceive.
\Fig{fig:intuition} plots the iso-chrome line for Deuteranopia\footnote{The line connects the spectral colors at the 475~$nm$ and 575~$nm$ in the xy-diagram.
More rigorously, these are actually two iso-chrome line segments that connect at [1/3, 1/3] in the xy-diagram.
The two line segments are almost parallel, so we treat them as if they are part of the same line to simplify exposition.}.
Naturally, the intersection of a confusion line and the iso-chrome line represents the color that a dichromat actually perceives when presented with all other colors on the confusion line.
For instance in \Fig{fig:intuition}, a \d sees  color $\alpha$ when presented with colors A, B, C. To make the ensuing discussion easier to follow, we use established trichromatic terminology for describing color shifts and visualize them in the 2D chromaticity diagram even though the perceived colors for a dichromat move only along the isochrome line.

\paragraph{Intuition.}
Consider two cases.
First, two confusing colors are on the opposite side of the iso-chrome line, e.g., color A (green) and color B (magenta) in \Fig{fig:intuition}.
They are initially seen by a \d as the same color $\alpha$ (a saturated yellow).
As the two colors are rotated in the RGB space, they go through the trajectories in the xy space as annotated by the corresponding dashed arrows.

Critically, from a \d's perspective, the green color at A shifts from $\alpha$ to $\beta$, transitioning from a desaturated yellow to a more saturated yellow;
in contrast, the magenta color at B shifts from $\alpha$ to $\delta$, transitioning from the desaturated yellow to a saturated blue.
That is, when the two confusing colors are on the opposite side of the isochrome lines, they shift in opposite directions.

Second, consider when two confusing colors are on the same side of the iso-chrome line, e.g., the magenta color B and the orange color C in \Fig{fig:intuition}
They will shift in the same direction but differ in the magnitude of shifts.
As we rotate B and C in the RGB space, they both shift toward the blue-ish hue, but it takes C more rotation to start appearing blue-ish (like color $\gamma$) than color B.
One can recognize the two colors by learning the amount of rotation needed for B and C to start becoming blue-ish or, alternatively, what the two colors look like after a particular amount of rotation.

Taken together, the rotational shifts in principle give rise to a unique third dimension: confusing colors on the opposite side of the iso-chrome lines shift in opposite directions, and confusing colors on the same side of the iso-chrome lines shift in the same direction but differ in the shift magnitude.
Critically, the shift of \textit{each} color is systematic: each color is cycled through the hue circle in a regular pattern.
We thus hypothesize that CVD individuals, through training, could learn to build a mapping from a particular color shift pattern to the color name (e.g., a shift from $\alpha$ to $\beta$ means green-ish color), thereby recognizing colors.
The rest of the paper provides empirical evidence to support this hypothesis through psychophysics and user studies.

\subsection{Design Decisions and Implementation}
\label{sec:idea:impl}

The user interface is implemented as a Web App.
The main rotation logic is implemented using JavaScript, which provides access to the smartphone camera.
No camera data is stored or transmitted outside the phone for user privacy.

\paragraph{Gesture.}
Users induce color shift is through the swipe gesture, which has one degree of freedom that corresponds to the rotation angle.
We have also considered other interfaces such as using the inherent 3 Degrees-of-Freedom (DoF) pose of the camera (yaw, pitch, roll), but neither works as well as the swipe gesture, which our participants suggest are easy to use (see \Sect{sec:real:painting} last paragraph).

\paragraph{Rotation Slider.}
For colors on the same side of the iso-chrome lines, recognizing colors relies on learning the amount of shifts  associated with different colors, which might not be intuitive to learn.
To assist learning, we add a slider at the top of the screen (see the top-left panel in \Fig{fig:demo}) to indicate the amount of shift being applied (i.e., rotation angle).
The slider helps users learn what a color looks like at different rotation angles, which are used to recognize colors.
We will later show at the end of \Sect{sec:long:disc} that this is largely the strategy that participants tend to use.

\begin{figure*}[t]
    \centering
    \subfloat[A trial in the discrimination task.]{
      \label{fig:pilot_setup}
      \includegraphics[height=2.in]{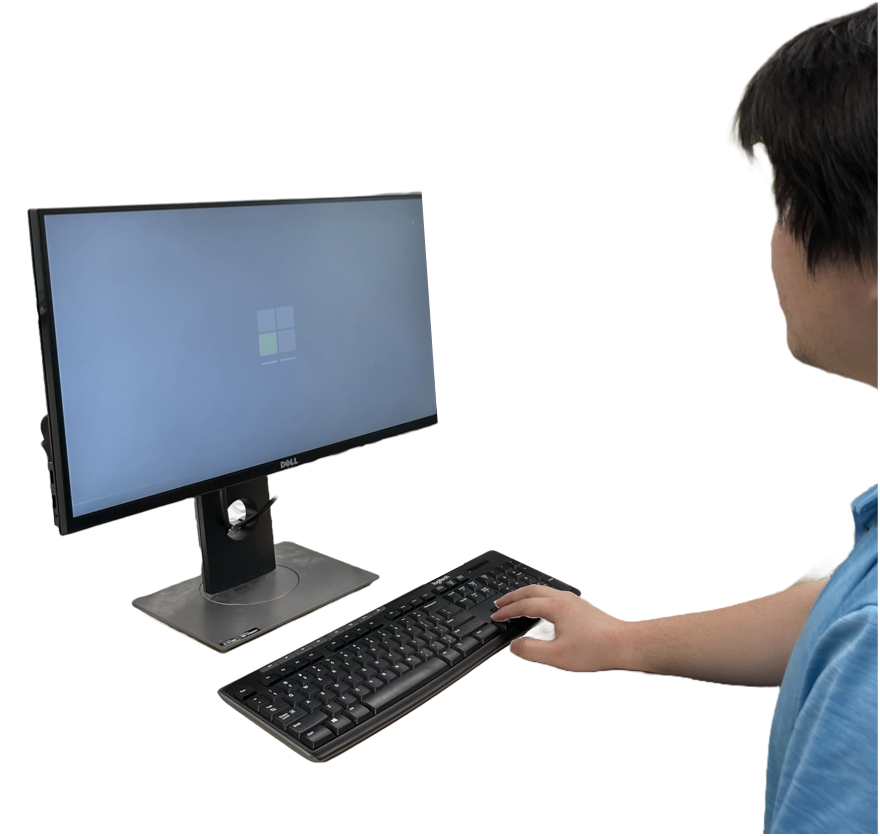}
    }
    \hspace{5pt}
    \subfloat[An example of the staircase procedure.]{
      \label{fig:staircase_new}
      \includegraphics[height=2.in]{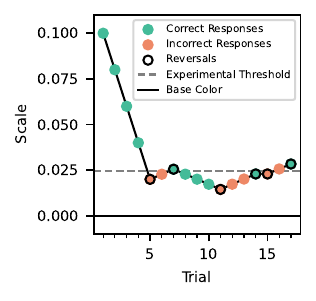}
    }
     \hspace{5pt}
    \subfloat[Discrimination thresholds of a base color.]{
      \label{fig:pilot_example}
      \includegraphics[height=2.in]{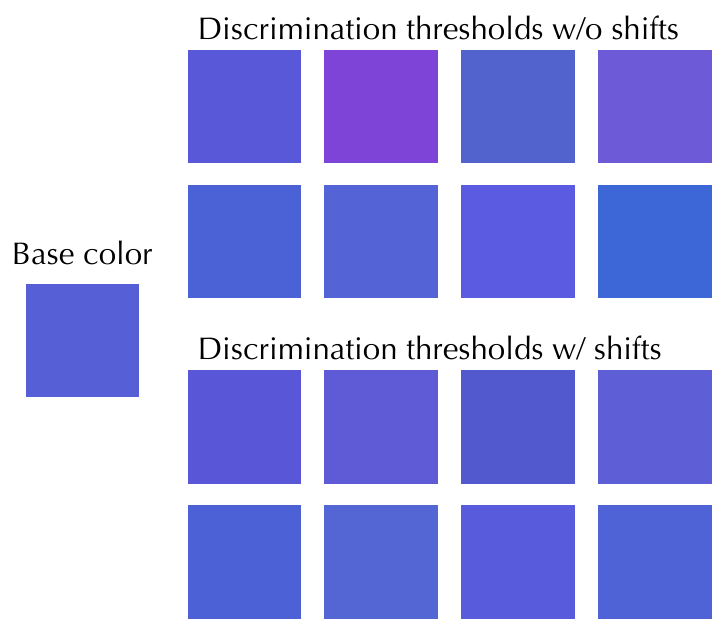}
    }
    \caption{\protect\subref{fig:pilot_setup} A trial in the discrimination task.
  There are four patches, three of which have the same base color and a randomly placed patch has the odd color. The participant is asked to identify the odd color (4AFC).
  In the first half of the study, they use the left/right arrow keys to induce color shifts, which are not available in the second half of the study.
  \protect\subref{fig:staircase_new} An example of the 1-up-2-down staircase procedure to narrow down the color-discrimination threshold.
  As a participant completes a 4AFC sequence in \protect\subref{fig:pilot_setup}, the staircase procedure automatically adjusts the odd color to make it harder if the participant identifies the odd color correctly and vice-versa.
  The $y$-axis is a relatively measure of the distance between the current color and the base color; 0 means the base color.
  \protect\subref{fig:pilot_example} For a given base color (left), we experimentally obtain eight discrimination thresholds without using the shifts (top) and eight thresholds while using the shifts (bottom).
    }
    \label{fig:pp}
\end{figure*}

\section{Psychophysics: Rotational Shifts Have Discriminative Power}
\label{sec:pp}

We perform a psychophysical study to show that rotation has discriminative power:
the color discrimination thresholds significantly reduce when rotational shifts are used.
We first describe the task and procedure (\Sect{sec:pp:task}), followed by the results (\Sect{sec:pp:res}) and a quantitative explanation of the results (\Sect{sec:pp:disc}).

\subsection{Task and Procedure}
\label{sec:pp:task}

\paragraph{Task.}
The goal of our study is to compare the color discrimination thresholds with and without using the color shifts.
To determine the discrimination thresholds, participants perform a series of four-alternative forced choice (4AFC) experiments.
\Fig{fig:pilot_setup} illustrates the task \change{using one trial as an example, where} the participant is presented with four color patches.
Three of the patches have the same ``base color'', which does not change throughout a sequence of trials.
The other patch, which is randomly placed, has a different color, which we call the ``odd color''.
The participant's task is to identify and select the patch with the odd color (using the keyboard).

To familiarize the participants with the user interface and minimize typos, we provide a training phase, in which the base color and the odd color are immediately distinct for the participants so that they can focus on getting familiar with the UI.
A participant must successfully identify the odd color six times in a row to pass the training before entering the tests.

\begin{figure*}
  \begin{minipage}[c]{0.25\textwidth}
    \caption{
       Color discrimination results averaged across Deuteranomaly participants (N=8).
       We have four base colors, each of which has two sets of discrimination thresholds, one without using the color shifts and the other using the shifts.
       For each set of thresholds, we regress an ellipse inspired by the classic MacAdam's ellipses~\cite{macadam1942visual} for modeling color discrimination thresholds.
       The discrimination threshold reduction is statistically significant for green, gray, and blue base colors.
    }
    \label{fig:ellipse_c}
  \end{minipage}  
  \hfill
  \begin{minipage}[c]{0.73\textwidth}
    \includegraphics[width=\textwidth]{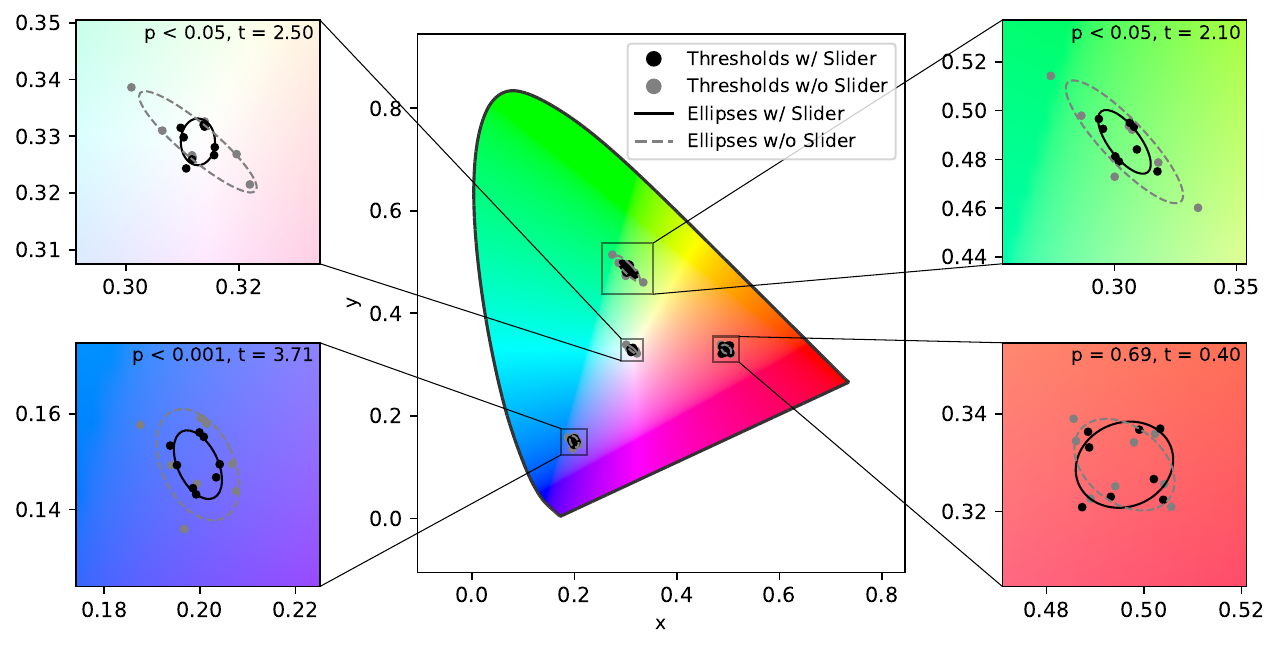}
  \end{minipage}
\end{figure*}

\paragraph{Stimuli.}
We choose four base colors, which can be generally described as blue (sRGB [86, 95, 214]), green (sRGB [100, 204, 102]), red (sRGB [184, 74, 74]), and gray (sRGB [136, 136, 136]).
They cover the primary colors of a RGB space along with the achromatic color, and are reported by a previous study as colors that CVD individuals commonly confuse with other colors~\cite{Flatla:ColourID:CHI2015}.
\change{\Fig{fig:ellipse_c} shows the locations of the four base colors on the CIE 1931 xy chromaticity diagram.}

To get the discrimination thresholds of each base color, we sample along four lines that roughly cover a full circle: the three confusion line directions of the Protanopia, Deuteranopia, and Tritanopia, along with the line orthogonal to the Protanopia confusion line.
Each trial has a randomly chosen base color and a randomly chosen discrimination direction (from among the two directions per line).
All the color patches are placed at the center of the display with a neutral gray background (i.e., [0.5, 0.5, 0.5] in the linear sRGB space) following prior work~\cite{duinkharjav2022color}.

\paragraph{Procedure.}
In each trial, depending on their answer, the odd patch's color was made easier or harder to discriminate in the subsequent trial by adjusting the odd patch along the line to be closer/farther from the base color.
The adjustment follows the classic 1-up-2-down staircase procedure commonly used in discrimination tasks in psychophysics~\cite{treutwein1995adaptive, garcia1998forced, lu2013visual}.
An example sequence under the staircase procedure is shown in \Fig{fig:staircase_new}.
After 6 reversals of the staircase procedure, the sequence terminates, and the discrimination threshold (color that is indistinguishable from the base color) is calculated as the average of the last three reversals.
The stimuli are displayed until an answer is picked.

Each study is split into two sections.
In the first section, participants are asked to use the left/right arrow keys to induce color shifts (to all four color patches), whereas in the second section the shift is not available.
The base colors are randomized within each section.
That way, we derive color discrimination thresholds both with and without using color shifts.
The difference between the two reveals the discrimination power of our rotational color shifts.
\Fig{fig:pilot_example} shows the results of one base color, where it can be seen that the eight discrimination thresholds without using shifts (top) appear visually more different from the base color that do the eight thresholds derived when colors shifts are induced (bottom).

In total, each study consists of 64 sequences (4 base colors $\times$ 4 sampling lines $\times$ 2 directions per line $\times$ 2 phases), and takes about 2 hours to finish.
Each participant is asked to complete a study twice.
Participants are encouraged to take breaks in-between sequences.
Upon returning from a break, participants can go through the training phase to re-familiarize themselves with the UI.

\paragraph{Participants.}
We recruit 16 CVD individuals (2 Protanopia, 2 Protanomaly, 3 Deuteranopia, and 8 Deuteranomaly; 3 female).
Each subject is asked to go through the Pilestone color blind test~\cite{pilestonetest} to classify their CVD type. 50\% of the subjects were categorized as Deuteranomaly, which agrees with the statistic that Deuteranomaly is the most common type of CVD~\cite{birch2001diagnosis, Birch:GreenRedWorldStatistics:JOSAA2012}.
All studies in the paper, including this one, are approved by our Internal Review Board (IRB).
All participants’ data were de-identified.

\subsection{Results}
\label{sec:pp:res}

Our results show that the rotational shifts have discriminative power.
For simplicity, we use results on Deuteranomaly to elaborate, and summarize the results for other CVD types in the end.

Recall that, for each base color and each participant, we obtain two sets of discrimination thresholds, one with and one without leveraging the color shifts.
We average the discrimination thresholds in each case across all Deuteranomalous participants.
For each case, we then regress an ellipse that best fits the corresponding thresholds.
The elliptical regression is inspired by MacAdam's classic psychophysical experiments that model human color discrimination thresholds as ellipses~\cite{macadam1942visual}.

\Fig{fig:ellipse_c} shows the results for Deuteranomaly with four blow-ups, each showing details for a base color.
For three base colors, using the shift significantly reduces the discrimination thresholds.
To test the statistical significance of the effect of color shifts,
we regress ellipses for each participant at each base color.
We then compare the areas of the regressed ellipses in the shift and no shift cases.
The elliptical area difference is statistically significant for the blue, green, and gray base colors
($p < 0.05$; a one-tailed paired t-test \change{under the null hypothesis that the mean areas of the ellipses before and after using the slider are not different}).
The reason why the result at the red base color is insignificant warrants further investigation (e.g., imperfectly calibrated displays).

The reduction of discrimination thresholds is even more statistically significant for Deuteranopia at the blue, green, and gray base color, all with $p < 0.01$.
Participants with Protanomaly have statistically significant decreases in discrimination thresholds for green and gray ($p < 0.05$).
In general, however, due to a lack of participants that have Protanopia and Protanomaly, we could not claim strong statistical significance for the two cases:
we have only 2 participants each with Protanopia and Protanomaly;
for comparison, we have 8 participants with Deuteranomaly and 4 participants with Deuteranopia.

\subsection{Discussion}
\label{sec:pp:disc}

\begin{figure}[t]
  \centering
  \includegraphics[width=\columnwidth]{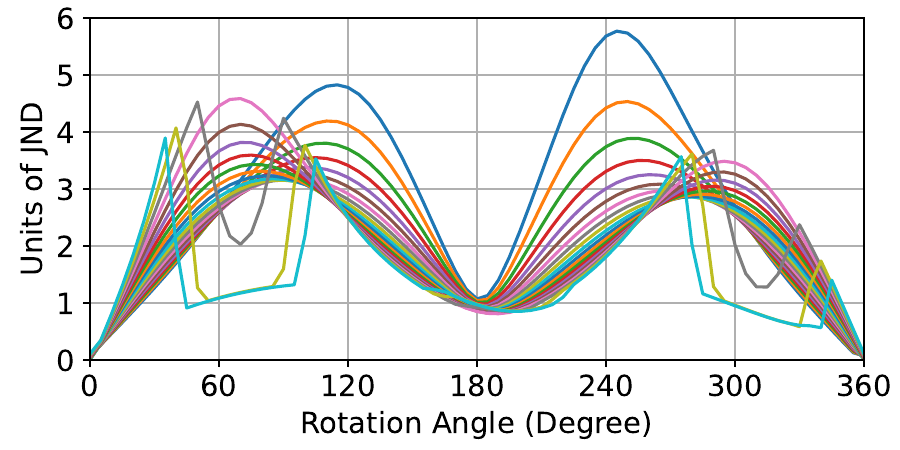}
  \caption{Each curve represents how the color difference of a pair of two colors changes with the rotation angle (degree).
  The color difference is plotted in units of Just Noticeable Difference (JND), where one JND is about 2.3 CIELAB $\Delta$E metric.
	Each pair consists of two colors that are about 2 units of JND apart on the gray base color's Protanopia confusion line.
	For \textit{each} pair, there exists an angle at which the difference between the two (initially confusing) colors is at least 3 JND, distinguishable for a dichromat.}
  \label{fig:diff}
\end{figure}

\begin{figure*}[t]
  \centering
  \includegraphics[width=\textwidth]{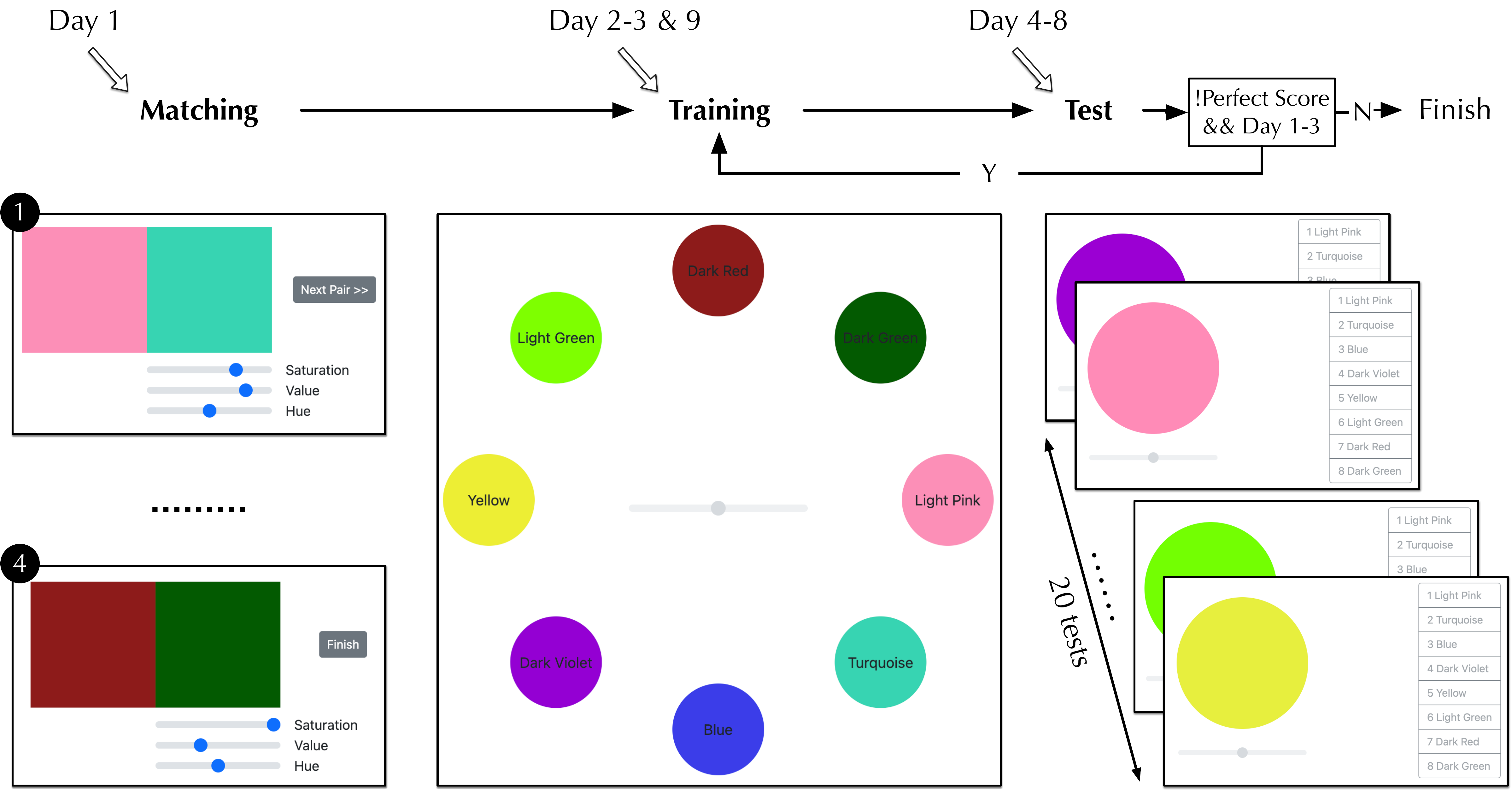}
  \caption{Procedure of the longitudinal color recognition study.
  See the last paragraph in \Sect{sec:bck} for a discussion of how we assign a name to a given color using a color naming dictionary familiar to trichromats.
  Even though each test question is an eight-alternative-forced-choice, participants indicated that they had no issues recognizing colors across pairs, so the only confusion is from colors within a pair, leading to an effective chance level of 10/20.}
  \label{fig:longitudinal-workflow}
\end{figure*}

It is not immediately obvious why the rotation shifts have discriminative power, given that CVD individuals can see only a small subset of colors, i.e., colors on the iso-chrome lines (see \Sect{sec:idea:why} and \Fig{fig:intuition}).
We provide a quantitative explanation.
The idea is to show that, during the rotational shift, there exists an angle at which the two confusing colors become different enough for CVD individuals to discriminate.

We show the gray base color results here as an example, and confirm the conclusion holds for others.
Along the \p confusion line for gray, we find a set of 13 colors that are about 5 CIELAB $\Delta$E away from each other, which represents about 2 units of Just Noticeable Difference (JND)~\cite{lu2013visual, gescheider2013psychophysics}.
We then form 12 pairs of colors, each consisting of two adjacent colors in the initial 13-color set.
As we rotate the colors, the difference between the two colors in each pair changes.
\Fig{fig:diff} plots, for each pair, how the color difference (plotted in units of JND) as seen by a \p ($y$-axis) changes as the rotation angle ($x$-axis) changes from 0$^{\circ}$ to 360$^{\circ}$.

During rotation, the \textit{maximum} color difference in any pair is at least 3 JND and can be up to 6 JND.
This means for \textit{each} pair of initially confusing colors (recall all the colors are from a confusion line), a dichromat can find an angle at which the difference between the two colors is clearly distinguishable
\footnote{While the CIELAB $\Delta$E color difference metric is designed for trichromats,
we argue that it applies well to \p{s} and \d{s}, because the confusion lines of Protanopia and Deuteranopia are almost orthogonal to the iso-chrome lines.
This means a CVD's ability to discriminate colors on the iso-chrome line (which, recall, contains all the colors that CVD individuals actually see) is only minimally affected by their color vision deficiency.}.

\section{Longitudinal User Study}
\label{sec:long}

Having demonstrated that rotational color shifts provide discriminative power, we conduct a longitudinal study to show that users can learn the colors shift patterns to recognize otherwise confusing colors.
Recall that discriminating two colors just means one can tell that the two colors are different, but our goal is stronger: we want CVD individuals to correctly name confusing colors.

\subsection{Study Overview}
\label{sec:long:ov}

\paragraph{Rationale.}
The goal of the study is to test the hypotheses that 1) CVD individuals can learn to associate temporal shifts with color names, 2) the learning is generalizable, even to unseen colors close to those seen in training, and 3) the learned association lasts days after the initial learning occurs.

To that end, we conduct a longitudinal study, where participants first train themselves to learn the shift patterns of four pairs of otherwise confusing colors; the shifts are voluntarily induced by the participants, who can strategize how to go about learning the patterns.
Over the next few days, the participants then are asked to use what they have learned from the training phase to name colors that they have not seen in training.

\paragraph{Participants.}
We recruit eight participants (1 Protanopia, 1 Protanomaly, 2 Deuteranopia, 4 Deuteranomaly; 3 female) from among the 16 participants for the color discrimination study mentioned in \Sect{sec:pp}.
The study has a Web interface, and participants complete the study using their own computers.

\subsection{Design}
\label{sec:long:proc}

\begin{figure*}[t]
    \centering
    {
      \label{fig:naming_legend}
      \includegraphics[height=1.8in]{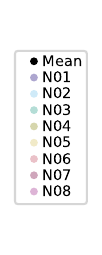}
    }
    \subfloat[Number of attempts during the first three days in the training phase.]{
      \label{fig:attempts_vs_day}
      \includegraphics[height=1.8in]{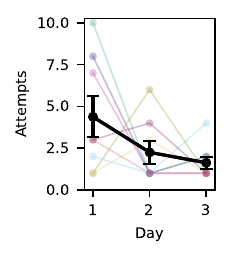}
    }
    \hspace{5pt}
    \subfloat[Test scores of eight participants and the average during the longitudinal study. Score at the chance level is 10.]{
      \label{fig:score_vs_day}
      \includegraphics[height=1.8in]{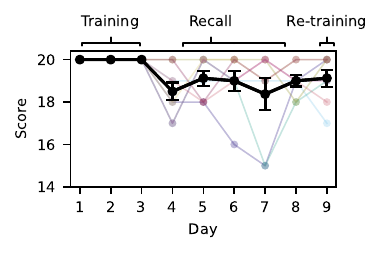}
    }
    \hspace{5pt}
    \subfloat[Average time spent on incorrect vs. correct answers.]{
      \label{fig:confidence_time}
      \includegraphics[height=1.8in]{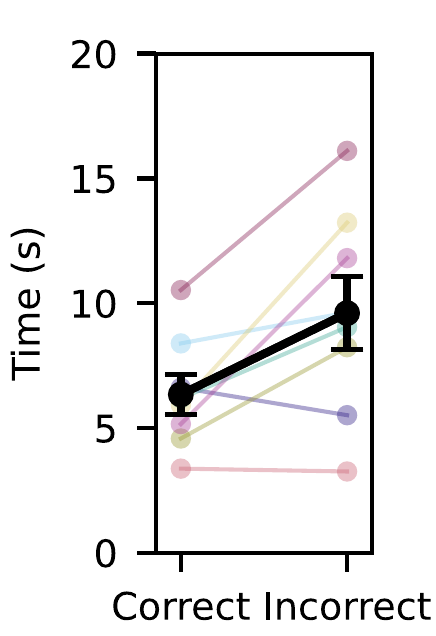}
    }
    \caption{Longitudinal study results.  In each figure, the thick line represents the mean across all eight participants, whose individual results are the semi-transparent lines in the background.  The error bars indicate Standard Error of the Mean (SEM).}
    \label{fig:naming_results}
\end{figure*}

The study spans 9 days and has four phases: color matching (Day 1), training (Day 1 -- Day 3), recall (Day 4 -- Day 8), and re-training (Day 9).
\Fig{fig:longitudinal-workflow} illustrates the procedure of the study that spans nine days.
On the first day, participants have to perform both color matching and training.
We now describe the design in detail.

\paragraph{Matching Phase (Day 1).}
Each participant is asked to first find four pairs of distinct colors that are confusing to them.
The left side panels in \Fig{fig:longitudinal-workflow} shows the matching interface.
For each pair, we fix the base color and ask the participant to adjust the other color for a match.
The matching color is initially set to be a color far away from the base color along the participant's confusion line.
We allow the participant to fine-tune the saturation, hue, and value of the matching color to find an identical match.
The fine-tuning is limited to the confusion line and its nearby region.
In the end, we make sure that the two matching colors are actual confusing colors rather than trichromatic matches.

Finding exact matches is critical; otherwise participants can rely on the initial difference to name colors.
Participants who could not find an identical match for each of the four color pairs discontinued the study and no data from such participants was considered further.

Since participants have different CVD types, it is entirely expected that the exact matches will differ across participants.
For instance, to match a pink color some participants used a turqoise-ish color and others used a gray-ish color.
Roughly speaking, the four matching color pairs are: dark red vs. dark green, light pink vs. turqoise/gray, dark violet vs. blue, and light green vs. yellow, as shown in the middle panel in \Fig{fig:longitudinal-workflow}.

\paragraph{Training Phase (Day 1 -- Day 3).}
In the first three days, participants first perform a training task with the goal of memorizing the associations between shift patterns and color names.
The user interface for the training phase is shown in the middle panel in \Fig{fig:longitudinal-workflow}, where all four pairs of colors (eight colors in total) are shown together, and colors from the same pair abut.
Each participant sees only four distinct colors, because colors from a pair are visually identical for the participant.
The participant use the left and right arrow keys on the keyboard to shift all the colors simultaneously, via the rotational color space transform.
Their goal is to learn to associate the shift patterns with color names.

To make sure participants are actively engaged in the training task, the participants are asked to take a test after training.
In the test, the participant is asked to name colors while using the shift voluntarily induced in the same way as in the training.
The test has 20 colors in total, repeating each of the 8 training colors (four pairs) twice, plus one randomly chosen color from each color pair.
The order of the 20 colors is randomized.
Participants are encouraged to spend as much as time they wish in training and take the test only when they feel comfortable.

In the training phase, the participants have to achieve a perfect test score, i.e., correctly name all 20 colors;
otherwise they will be prompted to go back to training and re-take the test until they score perfectly.
They can not immediately take another test.

To help participants generalize what they have learned to other similar colors, the test colors are perturbed from the training colors such that a test color is 4 units of CIELAB $\Delta$E away from the corresponding training color, which roughly corresponds to about 2 JND~\cite{lu2013visual, gescheider2013psychophysics} for color normal trichromats.
The underlying rationale is that a participant cannot simply memorize the shift pattern of a particular color during training; instead, they should find a strategy to learn underlying patterns of shifts that allow them to correctly name a similar color that is about 2 JND away from a training color and never directly seen during training.

\paragraph{Recall Phase (Day 4 -- Day 8).}
In the next five days, participants are asked to take the test (i.e., name 20 colors) daily \textit{without} having to score perfectly.
The 20 test colors are different each day, although they are still perturbed by about 2 JND from the training colors.
Upon completion of the test, they are told the correct answer to each question and shown their answers.

\paragraph{Re-training Phase (Day 9).}
On the last day, participants are asked to re-take the training (for as long as they want) and take the test again immediately after, without having to score perfectly.

\subsection{Results}
\label{sec:long:res}

\paragraph{Training Attempts.}
On the (three) training days, we recorded the number of attempts each participant needed to get a perfect score, and the results are shown in \Fig{fig:attempts_vs_day}, where the mean across the participants is also plotted.
Recall each attempt requires the participant to go through the training again before re-taking the test, and they are encouraged to take the test only when they are confident after each training (\Fig{fig:longitudinal-workflow}).

The average number of attempts required for the perfect score reduces from 4.375 on the first day to 2.25 and 1.625 on the second and third day, respectively.
The difference in number of attempts between the first training day and the last training day is statistically significant ($p < 0.05$, one-tailed paired T-test \change{under the null hypothesis that the mean number of attempts across participants on the first day and that on the last day are not different}), indicating that learning (of the association between the patterns of shifts and the color names) improves with repeated exposure to training.

\change{As a sanity check, we verified that the large number of attempts in the first training day is \textit{not} because one or two ``difficult'' pairs.
The three participants who had the largest number of attempts on the first day (N01, N03, N08) got only 10-12 colors (out of 20) correct on their first attempt and did not spend a majority of their time ``stuck on only one or two colors.''}

\paragraph{Recall Performance.}
\Fig{fig:score_vs_day} shows how the test score changes over the nine-day study for all eight participants and the average values.
The test scores are, by construction, perfect (20/20) on the first three days, since participants have to repetitively train and take the test until they get a perfect score (see \Fig{fig:longitudinal-workflow}).
Over the next five recall days, the average test score across participants is between 18.25 and 19.125.
The average scores on each of the recall days are well above the expected chance level (10/20), and the difference is statistically significant ($p < 0.01$, Z-test with $\mu_0 = 10$ \change{under the null hypothesis that the users’ scores would be no better than the chance level at 10}).

The results indicate that the memory of color shift patterns lasts days after initial training, suggesting that the temporal color shifts empower CVD individuals to recognize otherwise confusing colors significantly above chance level.

\paragraph{Re-training Performance.}
On the last day (Day 9), participants are asked to go through the training again and re-take the test.
The average test score on Day 9 is 19.125, which is not a significant improvement over the average score of the recall days ($p$ = 0.24, one-tailed T-test \change{underthe null hypothesis that the mean score across participants on the re-training day would be the same as that on all the recall days.}).
The result suggests that the color recognition performance tends to saturate even after re-exposure to the shift patterns.
This indicates that CVD individuals might need more specific learning guidance;
see our discussion in \Sect{sec:disc}.

\subsection{Discussion}
\label{sec:long:disc}

\paragraph{Confidence vs. Accuracy.}
\Fig{fig:confidence_time} compares the average time spent on tests that are incorrectly answered with that on tests that are correctly answered.
The former is statistically significantly longer than the latter ($p < 0.01$, one-tailed paired T-test  \change{with the null hypothesis that the the average time spent on incorrect answers and correct answers is the same}).
The time spent on a test question can be seen as a proxy of confidence: when participants hesitate more on a test they are more likely to get the answer wrong.

\begin{figure}[t]
  \centering
  \includegraphics[width=\columnwidth]{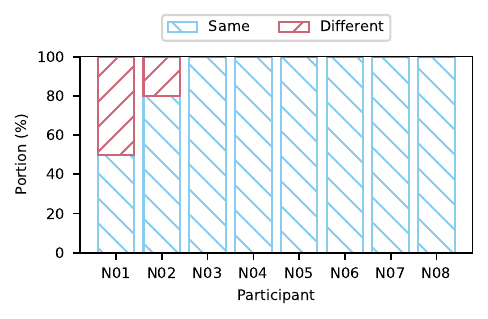}
  \caption{The distribution of incorrect test answers between two scenarios: colors that on the same side of the iso-chrome line and colors that are on the opposite side.}
  \label{fig:incorrect_ans_dist}
\end{figure}

\paragraph{Participants' Strategies.}
We conduct post-study interviews with the participants.
We first confirm the validity of the intuition we establish in \Sect{sec:idea:why}: across \textit{all} the participants, for colors on the opposite side of iso-chrome line they all report seeing colors shifting in different directions, and for colors on the same side of the iso-chrome line they all report seeing colors shifting in the same direction but with different magnitudes.

To learn colors shifting in opposite directions, they report using the strategy of observing how a color looks at different rotation angles with the assistance of the slider indicator.
This validates our design decision of adding a slider, as discussed in \Sect{sec:idea:impl}.
For instance, participants say ``\textit{The pink shifts much more than the turquoise which doesn't seem to shift much at all.}'' and ``\textit{for green vs. yellow if I moved the slider to the right, the yellow would be a darker more blue tone and the green would be a lighter more pink color.}''

\paragraph{Distribution of Incorrect Recognitions.}
Could a participant be more likely to get certain answers wrong?
\Fig{fig:incorrect_ans_dist} shows, for each participant over all the five recall days, the distribution of incorrectly answers between two cases: when test color belong to a pair that is on the same side of the iso-chrome line vs. on the opposite side.
Except N01, all other participants' incorrect test answers are heavily biased toward confusing colors that are on the same side of the confusion line.

The results suggest that confusing colors that are on the same side of the iso-chromes line are generally much harder to discriminate and recognize than those on the opposite side of the iso-chrome line.
This makes intuitive sense because, as shown in \Fig{fig:intuition}, colors on the opposite of the iso-chrome line shift in opposite directions (e.g., one becomes blue and the other become yellow), which is more easily learned than when the confusing colors are on the same side of the iso-chrome line, in which case the two colors shift in a similar pattern but differ in shift magnitudes, which even with the slider indicator could be more nuanced to observe and learn.

\change{\paragraph{Device Color Calibration.}
Even though different participants use different smartphones, each participant used their own device consistently throughout a study, which is representative of the actual deployment scenarios of our system. As a result, the device color calibration is controlled \textit{within} each participant's experience. Additionally, modern displays are well calibrated color wise.
For instance, a recent comprehensive benchmarking of modern smartphones shows that the color error is generally within 0.02 JND~\cite{displaycolorbenchmark}.}

\begin{figure*}[t]
  \centering
  \includegraphics[width=2.1\columnwidth]{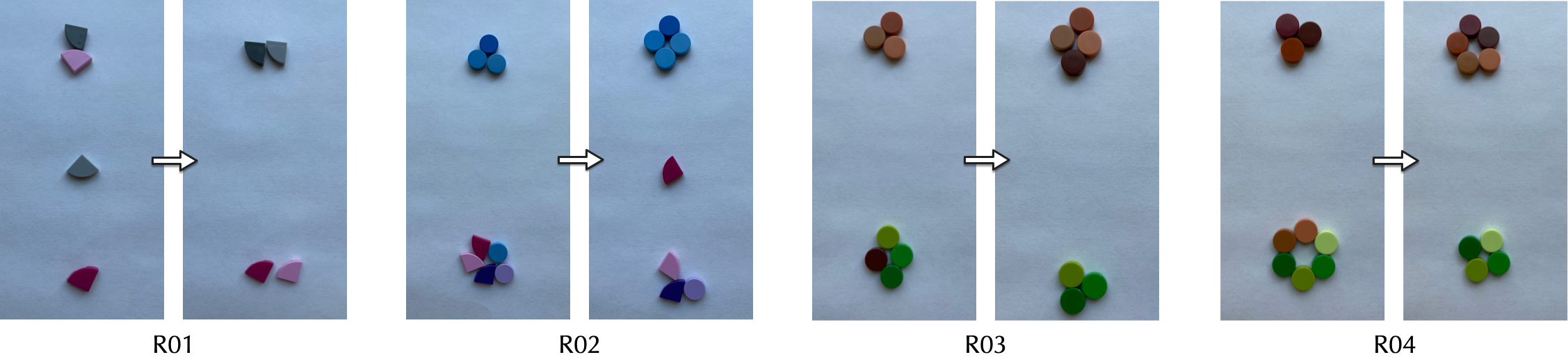}
  \caption{Results of sorting Lego blocks from four participants.  All participants initially had trouble correctly naming colors, but could successfully complete the task after using our App.}
  \label{fig:lego}
\end{figure*}

\section{Application Examples}
\label{sec:real}

We showcase the versatility of our proposed method with two real-world applications, where users use the mobile AR App in building with Lego blocks and observing AI-generated artistic images.
We invited six participants from the longitudinal study (1 female;  1 Protanopia, 1 Protanomaly, 2 Deuteranopia, 2 Deuteranomaly)\footnote{They are N01 -- N04, N06, and N08, which are renamed R01 -- R06 in this study.} to perform two essential steps in these tasks, namely recognizing colors of Lego blocks and perceiving colors in images.
The participants have all reported being familiar with the UI before the studies.

\paragraph{Rationale.}
Both tasks rule out context as a confounding factor.
It is known that CVD individuals sometimes can have a reasonable guess of an object color based on the context, e.g., an orange likely has an orange-ish color and the sky is likely blue.
In contrast, Lego blocks can be of any color, and AI-generated artistic images do not have to follow the actual object colors in the physical world.

It is also worth noting that \change{we do not get to control the exact colors of the Lego blocks and the colors in the artistic images;
these actual colors are generally not seen during the longitudinal study.}
Thus, performance on these two real-world scenarios is indicative of how users generalize what they have learned from training.

\begin{figure*}[t]
    \centering
    \subfloat[Secluded entrance to the ocean.]{
      \label{fig:entry_to_ocean}
      \includegraphics[height=4.6cm]{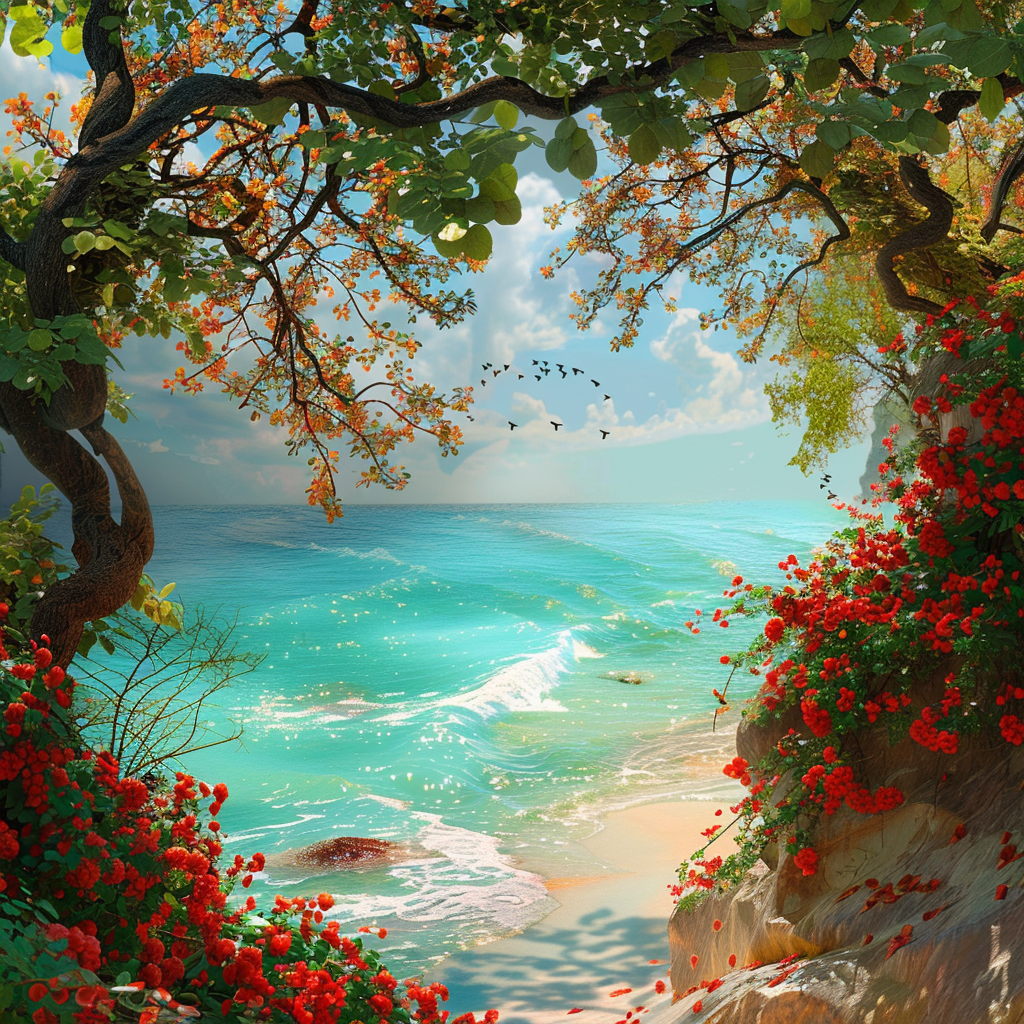}
    }
    \hfill
    \subfloat[Fall color.]{
      \label{fig:fall_color}
      \includegraphics[height=4.6cm]{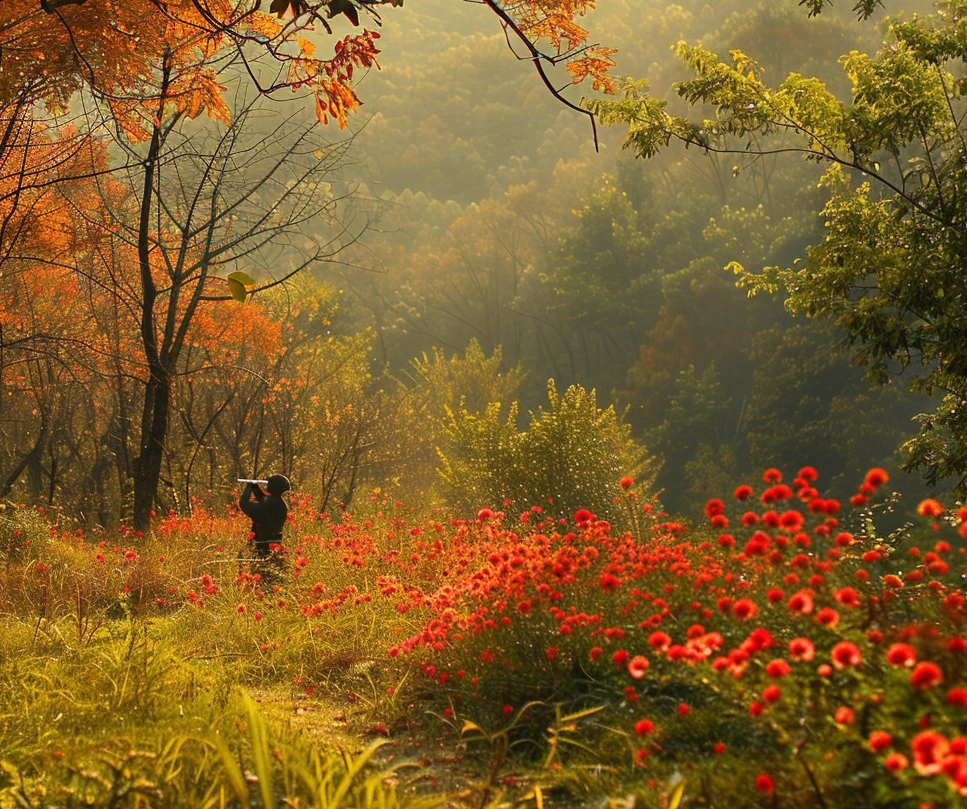}
    }
    \hfill
    \subfloat[Painting the ocean under cherry blossom.]{
      \label{fig:ocean}
      \includegraphics[height=4.6cm]{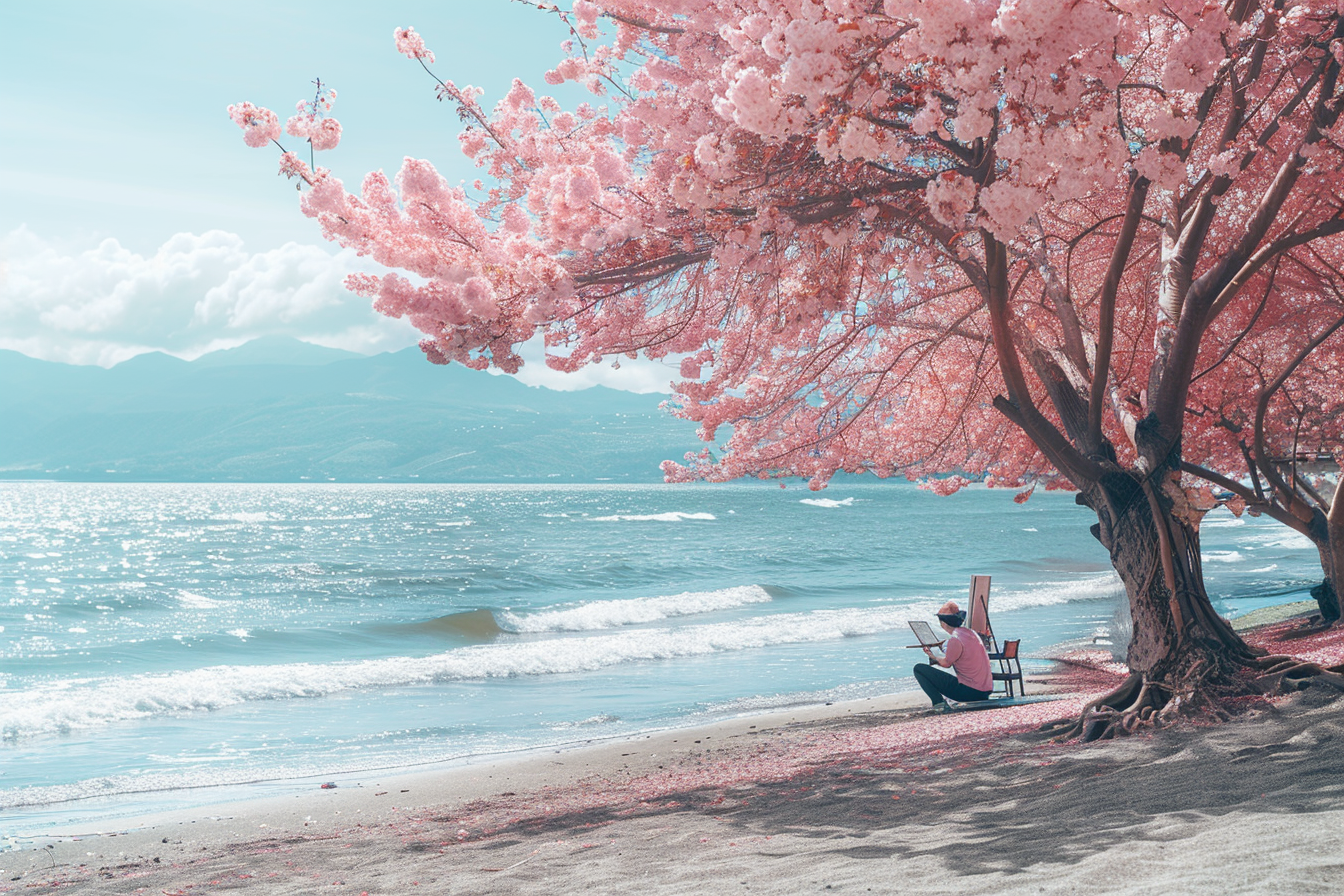}
    }
    \caption{Midjourney-generated images used by participants to recognize colors.
    }
    \label{fig:ai_examples}
\end{figure*}

\subsection{Building with Lego Blocks}
\label{sec:real:lego}

We choose building with Lego blocks as a representative task that requires users to physically manipulate individual objects with various colors.
In a building project, one necessary step is to identify a piece of block with the desired shape and color.
Our mobile App can assist users with CVD to accurately pick the intended blocks.
The users simply take a photo of the Lego blocks, as they perform the swipe gesture on the mobile phone, they can recognize the color of each Lego block and then select the desired ones to build up.

To verify the usability of our system in this task, we let the participants group Lego blocks by colors both using or without our App.
For each participant, \change{from the available choices,} we pick Lego blocks whose colors are likely confusing to them \change{based on} the matching results in the longitudinal study.
See \Fig{fig:lego} for examples
\change{\footnote{Specifically, the blue-purple pair is used by all participants; the green-brown pair is used by all but R02; the grey-pink pair is used by R01.}}.
We then tell the participant that the mix has multiple colors, each having different shades, which may or may not be confusing to them.
The participant is then asked to cluster the blocks according to their general colors and name each cluster.
The participant is asked to perform the task twice, one without using the App and the other with the App.

\paragraph{Results.}
\change{Only four of the six participants had challenges in this study.
With the App, they all completed the task successfully.
The other two were weak anomalous trichromats who could name the colors without the App.}
\Fig{fig:lego} shows the task result before (left) and after (right) using the App for the four participants.

We discuss the results of R01 as an example to showcase how our system helps address the color confusions.
Participant R01's mixed consists of two shades of pink (one lighter and one darker) and two shades of gray (again, one lighter and one darker).
R01 initially named the darker gray and the lighter pink both as gray-ish colors, the lighter gray as a cyan-ish color, and the dark pink (magenta) as a pink-ish color.
This is not surprising in that cyan, gray, and pink are confusing colors for Protanopes like R01, and without contexts R01 resorted to guessing.

With our AR app, R01 realized that the two gray-ish blocks did not shift much;
this suggested to him that both blocks were gray-ish colors, since gray colors lie on or close to the rotation axis and, thus, do not shift much during rotation.
In contrast, the other two blocks shift in patterns similar to what has been observed during training for pink, which suggests that they are pink-ish colors.

\subsection{Interpreting Artistic Works}
\label{sec:real:painting}

The second application is interpreting artistic works, which requires correct perception of the color patterns to fully comprehend an artwork.
To demonstrate the effectiveness of our system in assisting in this task, we show participants a set of images generated by Midjourney~\cite{midjourney}.
They are told that these images are AI-generated so object colors may or may not correspond to how they appear in the physical world.
The participants are first asked to describe the colors they see in the images without using the AR App, and are then asked to re-describe it using the App.

\paragraph{Results.}
\Fig{fig:ai_examples} shows the three artworks shown to participants.
\change{Each artwork is shown to all participants, but, due to their different types of CVDs, some artworks are more challenging for certain participants than others.
We now report the empirical findings from when participants use our App to explore each artwork.}

\Fig{fig:entry_to_ocean} shows an image that R01, R02, and R06 had challenges with.
Both R01 and R02 initially described the image having only red flowers at the bottom, and R06 described seeing only green flowers at the bottom.
They all quickly recognized green leaves intermixed with the red flowers at the bottom after using our App.

For the top part of the image they all initially saw only green leaves.
After applying the color shifts in the App, they all started seeing a different color.
R01 was still unsure what that color might be, R02 guessed that it had a yellow tint, and R06 guessed that it had a red tint.
In reality, the leaves at the top are a mix of green and orange colors.
No participant had seen orange-ish colors in their training, so the color shift might have appeared unfamiliar to them, leading to the hesitation.

Interestingly, however, R02 and R06 did correctly infer that the color had a tint of yellow and red, respectively, suggesting that they saw similarities between how the new color shifts and how yellow and red shift during training.
Indeed, orange is a mix of yellow and red; the three colors are close in the xy-chromaticity space and do have similar shift patterns (see \Fig{fig:rot_xy}).
This suggests that R02 and R06 were both able to generalize the learned patterns to qualitatively different colors.

\change{\Fig{fig:fall_color} was a challenging case for R03 and R06.
Prior to using the App, R03 described the image as having many red flowers on the right extending all the way to the left.
In reality, the middle and left parts of the photo have few red flowers but many green grasses, which R03 correctly recognized after using our App.
R03 explained that at first glance the image looked full of either green or red colors, which are confusing for the deuteranomalous vision.
However, the flowers on the right are easily recognized as flowers due to their shapes, based on which R03 guessed that the flowers are red since flowers are more likely red than green.
Since green and red appear virtually identical to them, without the App, R03 incorrectly assumed that the rest of the colors are all red-ish, too.}

\change{\Fig{fig:ocean} was a representative case for R01, a protanope, who, after using the AR App to view the image, was visibly surprised to realize that the colors in the image are very different from what he originally thought them to be.
Without using the App, R01 described the image as ``\textit{the ocean is red, the tree is pink, and the beach is gray if I have to take a guess.}''
R01 described the ocean as red presumably because the ocean is a cyan-ish color, which has a tint of green, which is confused with red for protanopes.
After using the App, R01 said the ocean is cyan and the beach is now definitely gray.}

\change{Prior to using the App, R04, who has protanomaly, described the ocean as pink, which makes sense as pink and cyan are confusing colors for protanomalous vision.
After using the App, R04 said the ocean is ``\textit{definitely not pink, but is most likely cyan or perhaps light blue}.''
This hesitation on light blue makes sense, because the ocean's cyan color, as a mix of green and blue, does have a tint of blue.
In our rotational shifts, blue and cyan do have similar shift patterns, since they are on the same side of the iso-chrome lines.
This suggests that R04 relied only on the shift pattern in recognizing colors while ignoring their different starting positions --- blue and cyan have different initial appearances even under the protanomalous vision.}

\subsection{Ease of Use}

After finishing both real-world tasks, each participant was asked to, anonymously, answer the question ``The smartphone App interface is easy to use'' using a 5-point Likert scale (1: strongly disagree; 3: neutral; 5: strongly agree).
\change{Our question is adapted from the System Usability Scale (SUS) question ``I thought the system was easy to use.'', which is the most relevant SUS question considering the relatively simple mobile App interface.}
The average rating is 4.0 (SD=0.58), indicating that our AR App is easy to use, and the result is statistically significant \change{($p$ < 0.01 using a one-tailed one-sample T-test against $\mu_0$=3, with the null hypothesis that the ease to use is no higher than neutral)}.

\section{Discussion}
\label{sec:disc}

\paragraph{Design Space of Temporal Modulation.}
The current work opts for rotational shifts as a computationally efficient and empirically powerful method for inducing temporal modulation.
The design space of temporal modulation, however, is vast.
In future work, we plan to explore the design space more comprehensively and characterize the pros and cons of alternative options, which we briefly describe below.

For instance, instead of rotating colors, which changes the hue, one could change the saturation and/or lightness of the colors.
We also do not have to apply the same temporal modulation uniformly to the entire visual field.
For instance, for a pair of confusing colors that are hard to recognize, we could apply a shift that intentionally amplifies the differences between the two colors during shift.
Such a heterogeneous mechanism can also be used to avoid hitting the gamut boundary during shifts.

In our current design, the speed of color rotation is proportional to the user's speed of swipe.
This works well because, in our observation, the swipe speed is roughly constant for a particular user, so the user can learn from consistent perceptual experiences.
In the future, we could explore an alternative where the swipe speed and the rotation speed are decoupled (i.e., a constant rotation speed regardless of the swipe speed), and understand how this design affects learning.
The decouple design could also potentially allow us to compare the effectiveness of learning across participants, since they all perceive the same rotation speed (for a given color).

\paragraph{Implementation on Headsets.}
While our AR application is implemented in a smartphone, an interesting future extension is to port it to an AR or a Mixed Reality headset, which would enable hands-free interactions for inducing the color shifts.
For instance, the colors can be rotated based on the virtual camera pose, e.g., as a user rotate the head or walks around the object in the scene.
With the hands free, users can also more freely use the gesture to control the shift.
For instance, we can design different hand gestures for different temporal modulation mechanisms, or allow users to freely draw, in the air, the shift trajectory in the RGB space.

\paragraph{Explicit Learning Guidance.}
An interesting direction for future exploration is to provide specific learning guidance to users.

From \Fig{fig:incorrect_ans_dist} and the discussion in \Sect{sec:long:disc}, we know that participants generally find it much harder to learn colors on the same side of the iso-chrome line, presumably because those colors have similar shift patterns and differ primarily in the speed with which they shift.
It is conceivable that ``speed'' is more nuanced to observe and to learn.
In the future, we could guide users to pay attention to the ``critical angle'' at which a color qualitatively changes (e.g., shifts to a different name for them).
In this way, instead of the different speeds, users learn to memorize the critical angles of the confusing colors, which could be more direct to learn.

In addition, we also observe that some participants pay attention only to the shift pattern but ignore the initial appearance of the color.
The analysis in \Sect{sec:idea:ui} shows that the initial appearance of a color provides the initial 2D percept needed to position the color in a 3D space and, thus, is critical for color recognition.
In the future, performance could be improved if we prime the participants with basic knowledge of color vision and CVDs.

\change{\paragraph{Automatically Applying Shifts.}
We considered the idea of automatically applying the color shifts without user control, but found it to be ineffective --- for three reasons.}

\change{First, the shift is context-dependent: there are many colors in a scene, and each user at any given time might care to name a specific color, which requires a specific shift.
Second, the shift is also user-dependent: different anomalous trichromats have different degrees of color deficiency, which means they need different shifts even for naming the same color.
For instance, the 4 deuteranomalous users' rotational angles can differ by over 30 degrees when naming Gray.
Finally, we also observe, and users' report, that they often need manual control of the slider to go back and forth, especially at the discrimination boundary, which is hard to do automatically.
That said, it would be a very interesting future work to explore how our system can learn from past interactions to infer the CVD degree/type, which could guide automatic shifts.}

\change{\paragraph{(Lack of) Comparison with Static Filters.}
We did not compare against approaches using static filters, which is the most commonly assistive technique for CVD individuals, as discussed in \Sect{sec:related:cd}.
Fundamentally, a static filter would allow only color \textit{discrimination} rather than color \textit{naming}, which is the focus of the paper.
Empirically, we also find that the way participants interact with our App indicates that a static filter does not help them name colors.
Depending on the colors they are asked to name, participants adjust the slider to different rotation angles, suggesting that a single filter is ineffective.}

\change{Our observation corroborates a recent empirical study on static color filters by Geddes et al.~\cite{geddes202330}, which points out that static filters have limited appreciation from actual color deficient individuals, because existing filters apply an one-off transformation whereas users often want to ``use information from all states (on/off and different recolouring tool modes) to make informed opinions on how they believed colours were transformed during recolouring.''
The continuous transformation afforded by our system provides such rich information.}

\paragraph{Larger Geographically Distributed Study.}
For color assistive technologies to be actually useful and address a real need, CVD individuals they seek to help should be closely involved in their development, as emphatically argued by Geddes et al.~\cite{geddes202330}.
In adherence with this philosophy, we closely engaged two CVD individuals (other than the participants who participated in the evaluation tasks), who participated in our weekly meetings and
provided feedback/suggestions via open-ended dialogues.

We chose a local cohort of CVD individuals as participants for our studies.
These studies have validated our scientific hypothesis and overall system design.
As a follow-up, we plan to significantly scale-up our study to engage a larger group of CVD individuals across different geographically regions to assess how our system can help them in day-to-day tasks involving color recognition.
This is enabled by our Web-based interface, which  users can access via their smartphones/laptops.

\section{Conclusion}
\label{sec:conc}

We propose, develop, and demonstrate a method that allow dichromatic individuals, whose color vision is ordinarily two-dimensional, to recognize/name otherwise confusing colors.
The key is to induce a new dimension through temporal color shifts, which, along with the initial two-dimensional percept of a color, positions colors in a new 3D space where users can learn to recognize and name colors.
Through psychophysics and a longitudinal study we show that the temporal color shifts have discriminative power and induce regular patterns that can be learned by users.
We implement an AR interface on smartphones.
The AR application is easy to use and helps CVD individuals recognize colors in two real-world scenarios (building with lego blocks and interpreting artistic works).

\section{Acknowledgement}
\label{sec:ack}

The research is partially funded by the Goergen Institute for Data Science (GIDS) Seed Funding Program at University of Rochester and NSF Award \#2225860.
We also thank Avery Young, Raj Rajwade, and Shreyan Goswami for participating in the early discussions of the project, and Suumil Roy for helping us set up the Web server for data collection.

\bibliographystyle{ACM-Reference-Format}
\bibliography{refs}

\end{document}